\documentclass{aa}

\usepackage{graphicx}
\usepackage{placeins}
\usepackage{txfonts}
\usepackage[table]{xcolor}
\usepackage{multirow}
\usepackage{enumitem}
\usepackage{hyperref}

\begin{document}

   \title{Peculiarities in the infrared emission of PAH-C$_{60}$ adducts}

   \author{R. Barzaga
          \inst{1}
          \and
          B. Kerkeni\inst{2}\fnmsep\inst{3}\fnmsep\inst{4}\fnmsep\inst{5}
          \and
          D. A. García-Hernández\inst{6}\fnmsep\inst{7}    \and
          X. Ribas\inst{7}
          \and
          T. Pelachs\inst{8}
          \and
          M. Manteiga\inst{9}
          \and
          A. Manchado\inst{2}\fnmsep\inst{3}\fnmsep\inst{10}
          \and
          M. A. Gómez-Muñoz\inst{11}\fnmsep\inst{12}
          \and
          T. Huertas-Roldán\inst{2}\fnmsep\inst{3}
          \and
          G. Ouerfelli\inst{13}
          }

   \institute{Departamento de Física and IUdEA, Universidad de La Laguna (ULL), E-38200 Tenerife, Spain \\
         \email{rbarzaga@ull.edu.es}
         \and
            De Vinci Higher Education, De Vinci Research Center,92 916 Paris, France
         \and
             Institut Supérieur des Arts Multimédia de la Manouba, Université de la Manouba, 2010 la Manouba, Tunisia
         \and 
             Faculté des Sciences de Tunis, Département de Physique, (LPMC), Université de Tunis El Manar, 2092 Tunis, Tunisia
         \and
             Sorbonne Université, Observatoire de Paris, Université PSL, CNRS, LERMA, F-92195 Meudon, France
        \and  
            Instituto de Astrof\'{\i}sica de Canarias, C/ Via L\'actea s/n, E-38205 La Laguna, Spain
         \and
             Departamento de Astrof\'{\i}sica, Universidad de La Laguna (ULL), E-38206 La Laguna, Spain
          \and
             Institut de Quimica Computacional i Catalisi (IQCC), Departament de Quimica, Universitat de Girona, Girona E-17003, Catalonia, Spain
          \and
             CIGUS CITIC - Department of Nautical Sciences and Marine Engineering, University of A Coruña, Paseo de Ronda 51, E-15011 A Coruña, Spain
          \and
             Consejo Superior de Investigaciones Cient\'{\i}ficas (CSIC), Spain
          \and
           Departament de Física Quàntica i Astrofísica (FQA), Universitat de Barcelona (UB), c. Martí i Franqués, 1, 08028 Barcelona, Spain
           \and
           Institut de Ciències del Cosmos (ICCUB), Universitat de Barcelona (UB), c. Martí i Franqués, 1, 08028 Barcelona, Spain
          \and
             Department of Physics, College of Khurma University, Taif University, P.O. Box 11099, Taif 21944, Saudi Arabia
            }

   \date{Received ??; Accepted ??}

  \abstract
  {The coexistence of polycyclic aromatic hydrocarbons (PAHs) and the C$_{60}$ fullerene in different astrophysical environments can give rise to the formation of new complex species denoted as PAH-C$_{60}$ adducts, which may contribute to the infrared (IR) emission observed. These PAH-C$_{60}$ adducts have been previously reported experimentally due to the high reactivity between PAHs and C$_{60}$. From the astrophysical point of view, however, they have not been considered in detail yet. Here we have performed a combined experimental and theoretical study in order to characterize the IR spectra of PAH-C$_{60}$ adducts, including multiple adducts. By using new advanced experimental techniques, we have been able to synthesize some specific PAH-C$_{60}$ adduct isomers, and measured their IR spectra. These experimental data are used to correct their harmonic scaled spectra, as obtained from quantum-chemistry calculations performed at the density functional theory (DFT) level under the B3LYP-GD3/6-31+G(d) approach. This way, we simulate the IR ($\sim$3$-$25 $\mu$m) spectra of multiple PAH-C$_{60}$ adducts, composed by a different number of PAH units: mostly one or two units. In addition, the chemical kinetics data available in the literature are used to tentatively estimate the possible order of magnitude of the abundances of these PAH-C$_{60}$ adducts using the available observational data. Essentially, our results reveal a possible strong modification of the IR spectra when astronomically estimated abundances are considered. Several spectral peculiarities are observed, such as a broad $\sim$3.4-3.6 $\mu$m feature, and important modifications in the 6-10 and 12-16 $\mu$m spectral regions together with contributions to the C$_{60}$ features at 7.0 and 18.9 $\mu$m. Interestingly, these PAH-C$_{60}$ adducts lack aliphatic CH bonds, but they display IR features around 3.4 $\mu$m, challenging previous interpretations of this astronomical feature.}


   \maketitle

\section*{Introduction}
Fullerenes, like C$_{60}$, C$_{60}^+$ and C$_{70}$, are the biggest molecules detected in space so far; either by their infrared (IR) emission bands or their visible absorption bands \citep{Cami2010,Campbell2015}. Specially neutral C$_{60}$, which has been detected via its IR transitions mainly in circumstellar envelopes of evolved stars such as young planetary nebulae (PNe) \citep{Cami2010,Garcia-Hernandez2011b}, but also in diverse astrophysical environments like R Coronae Borealis (RCB) stars \citep{Garcia2011a}, reflection nebulae \citep{Sellgren2010}, young stellar objects \citep{Roberts2012}, the diffuse interstellar medium \citep[ISM,][]{Berne2017} and more recently, in regions related to proto-planetary disks \citep{Iglesias-Groth2019, Arun2023}. The spectral signature of C$_{60}$ is recognizable by its four strongest mid-IR emission bands at $\sim$7.0, 8.5, 17.4 and 18.9 $\mu$m. According to these four bands, thermal models can be used to predict the C$_{60}$ temperature in different astrophysical objects; specially in the case of circumstellar envelopes of evolved stars where it has been estimated that an average temperature of $\sim$300 K is representative of the C$_{60}$ IR emission \citep{Cami2011,garcia2012}. In many of these astrophysical environments C$_{60}$ is accompanied by other organic species, which potentially could react with C$_{60}$, forming more complex molecular systems or C$_{60}$ derivatives. In fact, \citet{Barzaga2023} show that the variation of the C$_{60}$ 17.4$\mu$m/18.9$\mu$m band ratio observed in different C$_{60}$-rich circumstellar envelopes (PNe and RCB stars) may imply that there are other species like metallofullerenes, among possibly others, contributing at these wavelengths. This reinforces the idea of the possible presence of C$_{60}$ derivatives in space environments. One of the most obvious candidate molecules to form C$_{60}$ derivatives are polycyclic aromatic hydrocarbons (PAHs), which are detected in conjunction with C$_{60}$ in very different fullerene-rich sources \citep[e.g.][]{Garcia2010,Sellgren2010,Otsuka2013,Arun2023}.

The PAHs have been largely assumed to be responsible for the discrete unidentified IR (UIR) emission bands at $\sim$3.3, 6.2, 7.7, 8.6 and 11.2 $\mu$m widely observed in the Universe; from our Solar System to old stars and very distant galaxies, among others \citep[see e.g.][for a recent review and the comprehensive study of the Orion Bar]{Peeters2021,Peeters2024,Chown2024}. According to astrophysical models, the UIR emission bands can be attributed to large dimension PAHs composed by $\sim$25-100 C atoms \citep[e.g.][]{allamandola1989,Puget1989,Chown2024}. Indeed, a family of large PAHs (neutrals and ions) have to be randomly combined in order to reproduce the UIR bands \citep[e.g. in terms of the broadness of the features,][]{Rosenberg2014}\footnote{Note that the PAH model has been proved to reproduce the UIR emission bands, but it is not the only explanation; e.g. a family of PAH model spectra can also, in some sources, fit the spectra of several other species, both organic and inorganic ones \citep[see e.g.][]{Zhang2015}.}. However, such large PAHs have not been unambiguously detected yet in space; neither in the ISM nor any other astrophysical environment. In contrast, small cyano-PAHs like cyanonaphtalene \citep{McGuire2021} as well as the pure PAH indene \citep{Burkhardt2021,Cernicharo2021} have been recently detected for the first time in space, towards the cold dark cloud TMC-1. These detections have been only possible by radioastronomy, combined with microwave spectroscopy measurements, being the first convincing evidences of the existence of small PAHs in astronomical environments\footnote{Very recently, cyano derivatives of the PAHs acenaphthylene (C$_{12}$H$_{8}$), pyrene (C$_{16}$H$_{10}$) and coronene (C$_{24}$H$_{12}$) have been radioastronomically detected in TMC-1, being the largest PAH derivatives presently detected in space \citep{cernicharo2024,wenzel2025a,Wenzel2025b}.}.

The C$_{60}$ and PAH molecules can easily react to form new hybrid species denoted as C$_{60}$ adducts\footnote{The definition of C$_{60}$ adducts implies the binding of another organic molecule and/or functional group, e.g. alkyl chains, alcohols, among others.}, which are obtained by well-known organic chemistry experimental procedures \citep[e.g.][]{SAROVA2004,Garcia-Hernandez2013,Dunk2013,Barham2018}. In particular, these methods have been employed to synthesize C$_{60}$ adducts with acenes (anthracene, tetracene, pentacene), as well as with indene. The formation reaction of such C$_{60}$ adducts occurs smoothly, yielding single to multiple PAHs attached to the C$_{60}$ cage \citep[see e.g.][]{He2010,Cataldo2014}. This provokes that a mixture of regioisomers\footnote{Regioisomers are molecules with the same chemical composition and functionality, but with different spatial arrangement.} is produced, which is a main drawback, especially for their IR spectra measurements. However, a recent study reports the use of a supramolecular mask strategy in order to perform regioselective synthesis of anthracene and pentacene C$_{60}$ bis-adducts \citep{PUJALS2022}. This novel strategy allows to pinpoint the selection of the desired PAH isomers, even for multiple addend PAH units to the carbon cage \citep[see][for more details]{PUJALS2022}. In short, all previous experimental studies clearly show the trend of C$_{60}$ to easily form adducts under the presence of these small PAHs, something that it is very likely to also occur in astrophysical environments under UV-shielding conditions such as evolved stars, where both types of organic ingredients  may coexist (see Sect.\ref{AR}).

Herein, we present a quantum-chemical study, supported by novel experimental data, of multiple C$_{60}$ adducts with the small PAHs indene, indenyl, anthracene, tetracene and pentacene. Mono-adducts models are reported for all these PAHs, while the bis- and tris-adducts models are only included for the small PAHs indene, indenyl and anthracene. Their IR spectra have been simulated in order to capture the structural effects of the multiple addend PAH units and their relation with the IR spectral features observed. For this purpose, reliable scaling factors have been previously obtained and validated using the advanced experimental IR spectra measurements available for these C$_{60}$ adducts. Several spectral peculiarities are seen in the simulated IR spectra of multiple PAH-C$_{60}$ adducts, which are made publicly available to the astrophysical community, especially to users of the James Webb Space Telescope (JWST), for potential comparisons with their astronomical observations.

\section*{Methods}
Quantum-chemistry models of PAH-C$_{60}$ adducts require an accurate description of the numerous electrons system composing these molecules. The approach selected in order to carry out these quantum-chemical calculations is mainly determined by the number of electrons in the system. The density functional theory (DFT) is the natural choice to model large molecules containing more than 100 electrons. The PAH-C$_{60}$ adducts presented here contain up to $\sim$650 electrons, which corresponds to the maximum PAH units of anthracene added to C$_{60}$. The large system size makes thus unfeasible the treatment of our models with more accurate quantum-chemical approaches other than DFT; e.g. \textit{ab-initio} methods.
\subsection*{Computational details}
The Gaussian16 code \citep{g16} has been used to perform all the DFT calculations, in conjunction with the double zeta 6-31+G(d) basis set \citep{Petersson1988,Petersson1991}. Similarly to previous works on other PAH-C$_{60}$ adducts, the B3LYP functional \citep{Stephens1994} has been chosen to describe the exchange and electronic correlation \citep{BEHESHTIAN2012,KHODAMHAZRATI2016,Bakouri2018}. Large aromatic molecules like PAHs and C$_{60}$ can exhibit strong long-range forces (London forces) when they interact with each other \citep{Ehrlich2013}. In order to account for such forces, the third-level empirical dispersion correction of Grimme (GD3) has been also included in the calculations \citep{Grimme2010}.
Each PAH-C$_{60}$ adduct, after full optimization, was verified to be a stationary point with non-imaginary vibrations to be considered as a minimum. A thermodynamic analysis for all the PAH-C$_{60}$ adducts was conducted at the temperature of 300 K, along with the corresponding zero-point energy correction.

The IR vibrational spectra of all species were obtained within the framework of the harmonic oscillator and, subsequently, the harmonic frequencies were adjusted by applying a triple-scaling factor scheme to account for anharmonicity, vibro-rotational couplings, etc. (see next Section). The IR intensity has been modeled by a Lorentzian function as peak profile and with a full-width at half-maximum (FWHM) of 0.02 $\mu$m. This approximately reproduces the average spectral resolution (R$\sim$1700) of the Mid-IR Instrument (MIRI) onboard the JWST, working in the $\sim$5-30 $\mu$m spectral range. One of our goals is to provide the theoretically simulated IR spectra of PAH-C$_{60}$ adducts to the astronomical community as an useful tool to interpret JWST observational data. 

\subsection*{Scaling factors}
The predicted harmonic frequencies from quantum-chemical calculations must be corrected with scaling factors to avoid the overestimation with respect to the experimental fundamentals \citep{Trujillo2022}. Generally, scaling factors are determined by a fitting procedure of the theoretical IR spectral data against the experimental ones \citep[see e.g.][]{Xu2023}. In our case, we lack experimental data for all PAH-C$_{60}$ adducts under study and the standard procedure for the scaling factors fitting is not possible. Thus, we have applied an extrapolation method on existing scaling factors, depending of the PAH type bonded to the C$_{60}$ cage and based on the equation:
\begin{equation}
\nu_i \simeq  \mathrm{SF}_a \cdot \omega_i
\label{eqn1}
\end{equation}
where $\nu_i$ and $\omega_i$ stand for experimental and theoretical frequencies, respectively, with the $i^{th}$ suffix indicating that they correspond to the same normal mode and $\mathrm{SF}_a$ is the scaling factor. The Eq. \ref{eqn1} above can be reformulated as:
\begin{equation}
\nu_i \simeq  \mathrm{SF}_a \cdot \omega_i = \mathrm{SF}_a^{'} \cdot \omega_i^{'}
\label{eqn2}
\end{equation}
where Eq. \ref{eqn2} describes the relation between harmonic frequencies obtained from different quantum-chemical calculations and is fulfilled if the two sets of harmonic frequencies ($\omega_i$, $\omega_i^{'}$) are obtained at the same level of theory, producing thus scaling factors ($\mathrm{SF}_a$,$\mathrm{SF}_a^{'}$) that achieve a good agreement with the experimental values ($\nu_i$). Formally, the accuracy of harmonic frequencies not only depends on the level of theory used, but also on the basis set applied in the description of the quantum-chemistry model \citep{Trujillo2022,Trujillo2022a}. However, recent theoretical studies demonstrate that the median error in the obtained harmonic frequencies with respect to their experimental fundamental vibrations depends more on the level of theory selected than on the basis set applied \citep{Trujillo2023}. On the other hand, the scaling factors can correct the harmonic frequencies by applying a global (single number) scaling factor or a frequency-range-specific (multiple numbers) scaling factor. Frequency-range-specific scaling factors exhibit a higher accuracy than their global counterparts, which is due to the frequency range division performed. The frequency range is usually divided into three different frequency regions, each one of them with its corresponding scaling factor \citep{Trujillo2023}.

\begin{figure}
    \centering
    \includegraphics[width = 8.7cm]{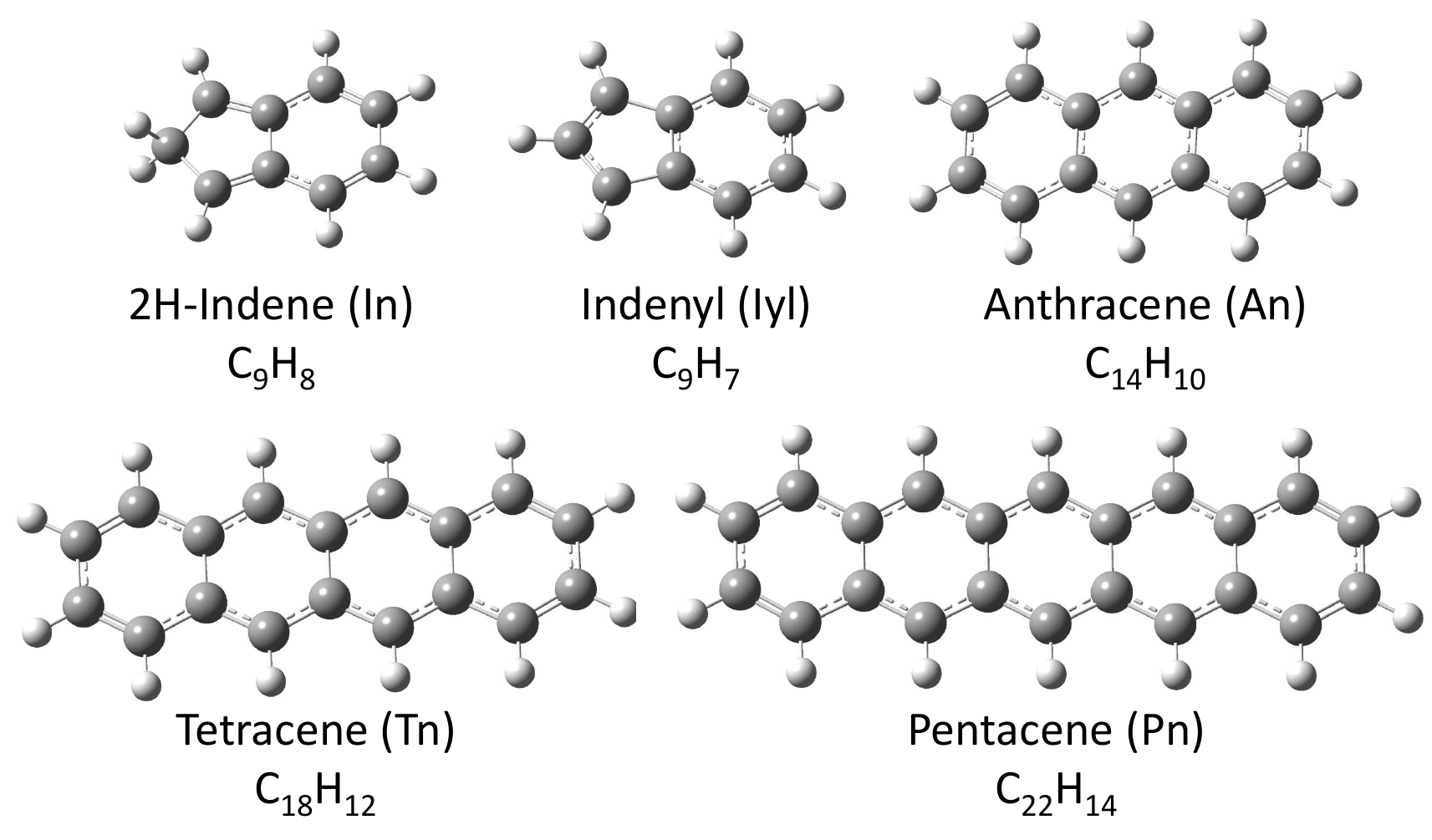}
    \caption{Structure of the several PAHs considered to build the PAH-C$_{60}$ adducts. The chemical formula, name and notations are also displayed.}
    \label{fig1}
\end{figure}
Following the concepts mentioned above to obtain new scaling factors for the PAH-C$_{60}$ adducts, we have applied Eq. \ref{eqn2} in combination with the frequency-range-specific scaling factors (at $<$ 5 $\mu \rm m$, 5-10 $\mu \rm m$ and $>$ 10 $\mu \rm m$) from the NASA Ames PAH IR Spectroscopic Database \citep[e.g.][AMES PAH Database hereafter]{Bauschlicher2018,Boersma2014,Mattioda2020}. The frequency-range-specific scaling factors of the AMES PAH Database have been computed at the 4-31G/B3LYP level for a large variety of PAH molecules, including the set of small PAHs of our interest (see Figure \ref{fig1})\footnote{For the particular case of indene, we only obtained the adduct formed by 2H-indene and C$_{60}$. According to our calculations, the 1H-indene, which is the most stable isomer, transforms into 2H-indene upon bonding to the C$_{60}$ cage.}. In addition to our standard 6-31+G(d)/B3LYP+GD3 approach, we have also computed the resulting harmonic frequencies at the 6-31+G(d)/B3LYP level only. The main purpose is to account for the effect of the dispersion correction (GD3) over the scaling factors.

\begin{table}
      \caption[]{Frequency-range-specific scaling factors for the PAHs in Figure \ref{fig1}.}
         \label{tb1}
         \begin{tabular}{llllll}
            \hline
            \noalign{\smallskip}
            PAH &  $\mathrm{SF_3^{AMES}}$ & $\mathrm{SF_3^{B3LYP}}$ & $\mathrm{\overset{B3LYP}{Max.r.e.}}$ & $\mathrm{SF_3^{D3}}$ &$\mathrm{\overset{D3}{Max.r.e.}}$\\
            \noalign{\smallskip}
            \hline
            \noalign{\smallskip}
                        & 0.9563\tablefootmark{a} & 0.9834 & 0.0132 & 0.9852 & 0.0269\\
             \text{In} &  0.9523\tablefootmark{b} & 0.9677 & 0.0124 & 0.9675 & 0.0151\\
                        & 0.9595\tablefootmark{c} & 0.9612 & 0.0025 & 0.9614 & 0.0023\\ \arrayrulecolor{gray} \cline{2-6}
             \noalign{\smallskip}
                        & 0.9563 & 0.9830 &  0.0099 &  0.9838  & 0.0093\\
             \text{Iyl} & 0.9523 & 0.9665 &  0.0074 &  0.9664  & 0.0114 \\
                        & 0.9595 & 0.9621 &  0.0026 &  0.9627  & 0.0028 \\ \cline{2-6}
              \noalign{\smallskip}
                        & 0.9794 & 0.9815 & 0.0109 & 0.9807 & 0.0190 \\
             \text{An} &   0.9691 & 0.9734 & 0.0103 & 0.9711 & 0.0114\\
                       & 0.9597 & 0.9603 & 0.0001 & 0.9610 & 0.0003\\ \cline{2-6}
              \noalign{\smallskip}
                       &  0.9563 & 0.9830  & 0.0103 & 0.9818 & 0.0118 \\
             \text{Tn} &  0.9523 & 0.9677  & 0.0120 & 0.9679 & 0.0037 \\
                       & 0.9595  & 0.9608  & 0.0012 & 0.9614 & 0.0009 \\ \cline{2-6}
              \noalign{\smallskip}
                       & 0.9563 & 0.9854 & 0.0402 & 0.9848 & 0.0500\\
             \text{Pn} & 0.9523 & 0.9676 & 0.0102 & 0.9648 & 0.0113  \\
                       & 0.9595 & 0.9612 & 0.0013 & 0.9618 & 0.0010  \\
            \arrayrulecolor{black}\hline
         \end{tabular}
      \tablefoot{The scaling factors from the AMES PAH Database (SF$_3^{\mathrm{AMES}}$) are compared to those from this work: 6-31+G(d)/B3LYP (SF$_3^{\mathrm{B3LYP}}$) and 6-31+G(d)/B3LYP+GD3 (SF$_3^{\mathrm{D3}}$). The maximum relative errors (Max.r.e) of the scaled frequencies are also included for both SF$_3^{\mathrm{B3LYP}}$ and SF$_3^{\mathrm{D3}}$. The specific frequency (or wavelength) range of harmonic frequencies for each scaling factor in this work ($\mathrm{SF_3^{B3LYP}},\mathrm{SF_3^{D3}}$) is according to the standard range set proposed by \citet{Trujillo2022,Trujillo2022a}. The wavelength ranges are highlighted for indene (In) as example: \tablefoottext{a}{$>$ 10 $\mu \rm m$}, \tablefoottext{b}{5-10 $\mu \rm m$}, \tablefoottext{c}{$<$ 5 $\mu \rm m$}. Max.r.e has been computed using wavelength units.}
   \end{table}
Table \ref{tb1} summarizes the comparison of frequency-range-specific scaling factors obtained through Eq. \ref{eqn2} with respect to those of the AMES PAH Database \citep{Bauschlicher2018,Boersma2014,Mattioda2020}. The new scaling factors $\mathrm{SF_3^{B3LYP}}$ and $\mathrm{SF_3^{D3}}$ provide the same level of accuracy, which is clearly reflected in the maximum relative error (Max.r.e.)\footnote{The Max.r.e is defined as the maximum value from the relative error of the new scaling factors with respect to the AMES PAH Database ones. The Max.r.e. is calculated using the scaled harmonic frequencies for each method according to the equation: $\rm Max.r.e<\frac{SF_3^{new}-SF_3^{AMES}}{SF_3^{AMES}}$.}, being very similar for both methods. The Max.r.e indicates the deviation of the scaled frequencies obtained from $\mathrm{SF_3^{B3LYP}}$ and $\mathrm{SF_3^{D3}}$ with respect to those of $\mathrm{SF_3^{AMES}}$. Our new scaling factors mostly have a maximum relative error of the order of $\sim$10$^{-2}$, with a slightly larger value ($\sim$5$\times$10$^{-2}$) at wavelengths longer than 10 $\mu$m (see Table \ref{tb1}). However, the spectral resolution of the experimental IR spectra used to obtain the AMES PAH Database scaling factors $\mathrm{SF_3^{AMES}}$ is significantly higher than our Max.r.e. values \citep[see e.g.][]{Mattioda2020}. So, the error introduced by the new scaling factors $\mathrm{SF_3^{B3LYP}}$ and $\mathrm{SF_3^{D3}}$ can be considered as negligible. In short, we conclude that both scaling factors $\mathrm{SF_3^{B3LYP}}$ and $\mathrm{SF_3^{D3}}$ can reproduce fairly well the vibrational frequencies (or IR features) of the small PAHs displayed in Figure \ref{fig1}. In the following, we only use the scaled frequencies as obtained from the new scaling factors $\mathrm{SF_3^{D3}}$.

\subsection*{Scaling factors validation}\label{SCF}
\begin{figure}
    \centering
    \includegraphics[width = 7cm]{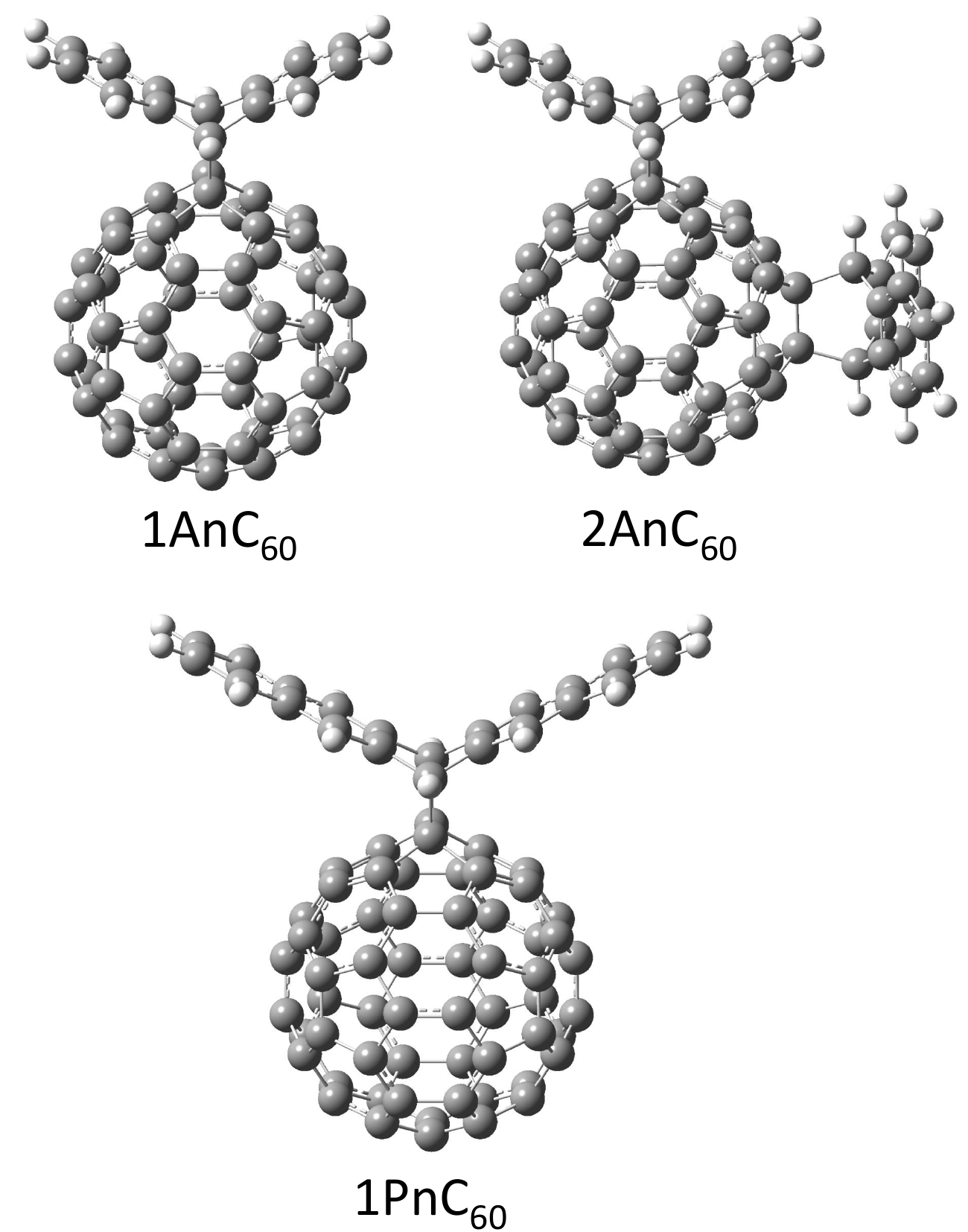}
    \caption{The three archetypal PAH-C$_{60}$ adducts synthesized by supramolecular mask strategy and used for the scaling factors validation: 1An$C_{60}$ (mono-anthracene), 2An$C_{60}$ (bis-anthracene) and 1Pn$C_{60}$ (mono-pentacene) \citep[see][for more details]{PUJALS2022}. The notation follows the legend in Figure \ref{fig1}, adding the number of PAH units and the C$_{60}$ identification.}
    \label{fig2}
\end{figure}

Despite the scaling factors $\mathrm{SF_3^{D3}}$ reliably reproduce the experimental IR features of small PAHs (Fig.\ref{fig1}), here we are interested in their adducts with the C$_{60}$ molecule. Therefore, it is not appropriate to directly extrapolate our $\mathrm{SF_3^{D3}}$ scaling factors to PAH-C$_{60}$ adducts before a proper validation. In order to validate our scaling factors, we have used the laboratory IR spectra measurements of three different types of PAH-C$_{60}$ adducts (see Figure \ref{fig2}). These three archetypal systems have been synthesized via a novel supramolecular mask strategy \citep{PUJALS2022}, which allows us to selectively obtain the bis-adducts (two PAH units) or mono-adducts (one PAH unit). It is worth noting that the synthesis route via supramolecular mask strategy can pinpoint to the desired PAH-C$_{60}$ adduct isomer. Thus, we are completely confident that the laboratory IR spectra are not contaminated by a mixture of regioisomers.

The experimental IR spectra of the three PAH-C$_{60}$ adducts in Figure \ref{fig2} were obtained by a Bruker ALPHA II FT-IR spectrometer with a spectral resolution of $\pm$2 cm$^{-1}$ equipped with a DTGS (deuterated triglycine sulfate) detector. All measurements were carried out in the solid state of the sample by using a Bruker Platinum ATR Adapter (attenuating total reflectance module with a diamond crystal support), allowing a direct measurement without prior treatment of the sample. All spectra were recorded at room temperature and under air atmosphere. The FT-IR measurements were recorded between 2.5 and 25 $\mu$m (4000-400 cm$^{-1}$). Each final spectrum was obtained by averaging 24 recorded scans.

In order to quantitatively assess the agreement between the experimental and theoretical IR spectra, we have followed the procedure of the cosine similarity score recently introduced in the literature \citep{Fu2018,Kempkes2019,Muller2020}. This similarity score method calculates the cosine of the angle $\theta$ between two n-dimensional vectors using their normalized Euclidean dot product according to:
\begin{equation}
    \rm{ Similarity=cos(\theta)=\frac{A\cdot B}{||A||\, ||B||}=\it{\frac{\sum_{i=1}^{n}A_{i}B_{i}}{\sqrt{\sum_{i=1}^{n}A_{i}^{2}}\cdot\sqrt{\sum_{i=1}^{n}B_{i}^{2}}}}}
\label{eq1}
\end{equation}
Where the elements $A_{i}$ and $B_{i}$ are the experimental (A) and theoretical (B) IR intensity at the same $i^{th}$ frequency. Concretely, the procedure is to compute the similarity score at the same reference frequency; both experimental and theoretical spectra have thus the same $x$-axis. In this case, the experimental spectrum is taken as reference and the similarity score was calculated at the exact experimental wavenumber. A similarity score close to 1 indicates a higher resemblance between the experimental and theoretical spectra.

On the other hand, \citet{Kempkes2019} introduce an alternative definition of the IR intensity to make the cosine similarity score more sensitive to the overlap of frequencies in the spectra $A$ and $B$ of Eq. \ref{eq1}, but diminishing the importance in the deviations of their IR intensities \citep[see][for more details]{Kempkes2019}. The \citet{Kempkes2019} IR intensity definition stand as:
\begin{equation}
    A_i^{rev} = log\left(\frac{A_i}{A_m} + \it{c}\right)
    \label{eqn3}
\end{equation}

\begin{figure*}
    \centering
    \includegraphics[width = 16.5cm]{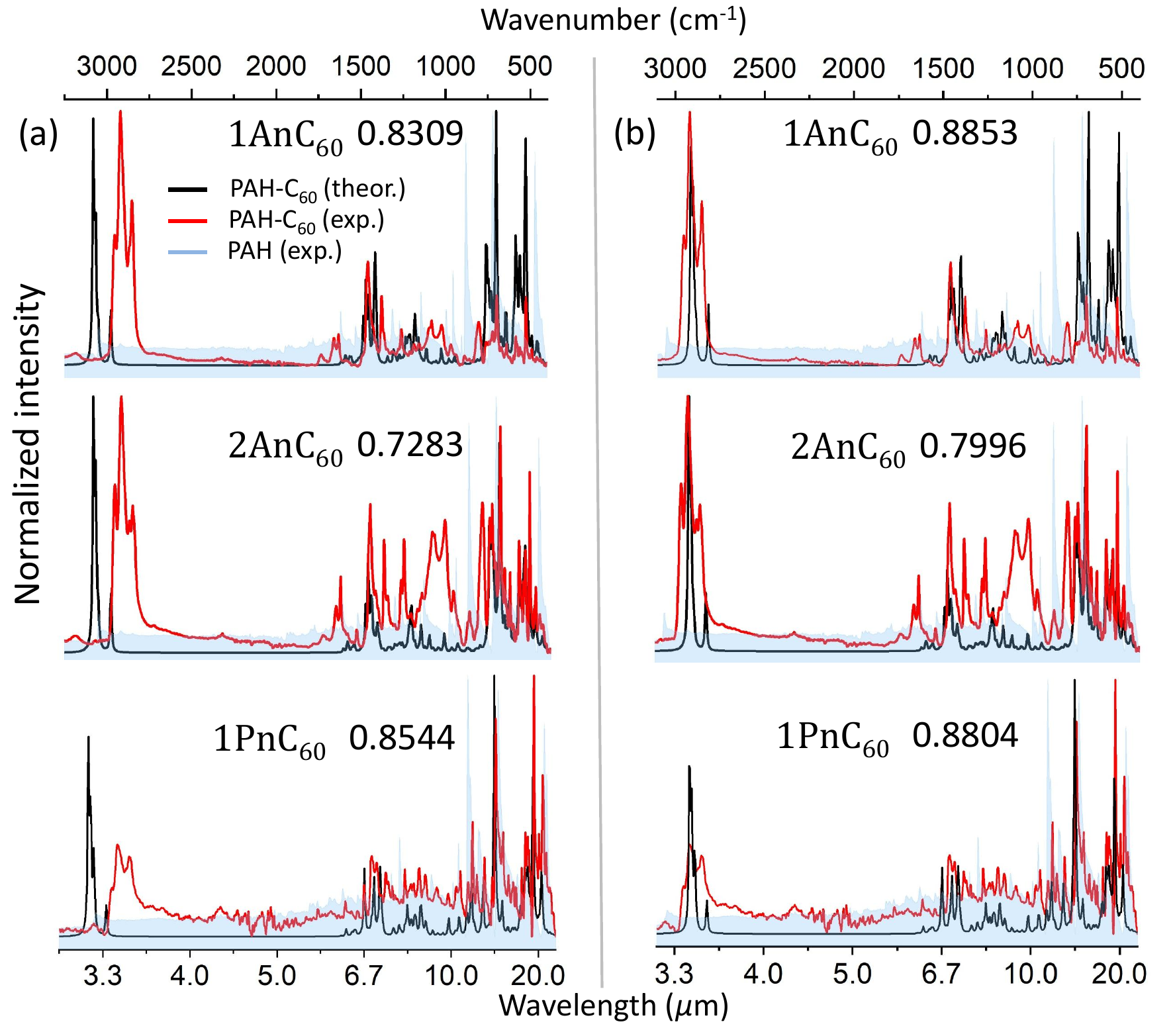}
    \caption{Experimental (red) versus scaled harmonic B3LYP-D3/6-31+G* (black) spectra for the three archetypal PAH-C$_{60}$ adducts used for the scaling factors validation: 1An$C_{60}$, 2An$C_{60}$ and 1Pn$C_{60}$ (Fig.\ref{fig2}). The experimental spectra of the pristine PAHs (light blue) are included to ease the discussion.  (A): Spectra built from scaling factors in Table \ref{tb1}. (B): Spectra built from empirical-corrected scaling factors for wavelengths < 5 $\mu$m ( > 2000 cm$^{-1}$) (see the text for more details). The similarity score value for each case is also indicated. The theoretical spectra have been constructed in the wavenumber scale by the convolution of a Lorentzian function with FWHM = 4 cm$^{-1}$, in order to ease the comparison with the experimental spectra.}
    \label{fig3}
\end{figure*}
The variable $A_i^{rev}$ denotes a new recalculated intensity, which is computed from the normalized intensity up to a maximum of 1 $\equiv$ ($ \frac{A_i}{A_m}$) and weighted by a factor $c$. We apply this procedure (Eq. \ref{eqn3}) to both the experimental (A) and theoretical (B) IR spectra, compromising the similarity score with the sensitivity to low-intensity bands in the spectrum, but avoiding the experimental noise. The factor $c$ is a constant for both A and B, which has to be derived from the correlation between the experimental and theoretical spectra. However, due to the lack of a significant number of resolved IR features (i.e. poor statistics) in our PAH-C$_{60}$ adducts experimental spectra (Fig.\ref{fig3}), we have not correlated them with the theoretical ones. Instead, we have used a factor of $c$ = 0.71 recently reported by \citet{Xu2023}, as obtained from a significant number of experimentally resolved IR features on smaller C$_{60}$ adducts like C$\rm_{60}O^+$ and C$\rm_{60}H^+$. This way, the recalculated experimental and theoretical intensities, $A_i^{rev}$ and $B_i^{rev}$, are used in Eq. \ref{eq1} to compute the similarity score values for the three archetypal PAH-C$_{60}$ adducts and indicated in the legend of Fig.\ref{fig3}. 

In Figure \ref{fig3}a we show that the scaled harmonic theoretical IR spectra of our three archetypal PAH-C$_{60}$ adducts are in good agreement with the experimental spectra regarding the vibrational frequencies. This is evidenced by the similarity scores obtained, which resemblance to those reported in previous experimental-theoretical studies \citep{Muller2020,Kempkes2019}. However, according to our experimental spectra, the archetypal PAH-C$_{60}$ adducts do not display IR features below $\sim$ 3.3 $\mu$m (above 3000 cm$^{-1}$). Apparently, our scaling factors produced from free PAHs cannot reproduce this spectral region for PAH-C$_{60}$ adducts. Thus, we have opted to apply an empirical correction to the scaling factors in Table \ref{tb1}, but only in the < 5 $\mu$m spectral range. The procedure imply a shift of the theoretical spectra until a maximum in the similarity score was obtained with respect to the experimental data. Figure \ref{fig3}b illustrates the empirical-corrected theoretical spectra this way, showing $\sim$ 3.3 $\mu$m region in better agreement with the experimental data. Using these three archetypal PAH-C$_{60}$ adducts we have found an empirical wavelength correction of 0.1984 $\mu$m, which was applied to the scaling factors for the rest of PAH-C$_{60}$ adducts. The corresponding scaling factors finally adopted can be found in Appendix \ref{SFemp}. It is worth noting that the experimental spectra of the pristine PAHs and PAH-C$_{60}$ adducts are strongly different in the $\sim$3.3-3.6 $\mu$m region (see Figure \ref{fig3}b); something that is discussed in the following sections.

However, deviations in the intensities are noticeable; specially in the case of 2AnC$_{60}$, which shows the largest intensity deviations of the three PAH-C$_{60}$ adducts (e.g. at wavelengths longer than 7 $\mu$m but also in the $\sim$3.3-3.5 $\mu$m spectral region). The difference in intensity between the theoretical and experimental IR spectra is likely due to several reasons. The collection of the experimental spectra is performed in solid state, which can modify the IR intensity with respect to gas-phase measurements like IR multiple-photon dissociation \citep[IRPMD,][]{Polfer2011,palotas2020} or even more sophisticated techniques \citep{Gerlich2018,Roithova2016}. In addition, we can not completely discard whether a better fitting of the experimental spectra baseline and/or a reduction of the noise contribution could improve the intensity match with the theoretical spectra. Nevertheless, we note that the experimental measurements of the three archetypal PAH-C$_{60}$ adducts are only used to validate our new derived scaling factors (see Table \ref{tb1} and Appendix \ref{SFemp}), which, according to the similarity score values, perform quite well regarding the vibrational IR frequencies. The intensity mismatch is a long-standing and well-known problem in the comparison of experimental and theoretical IR spectra of complex organic compounds \citep[see e.g.][]{Katsyuba2013}, and its resolution is, of course, beyond the scope of the present work.

\section*{Multiple PAH-C$_{60}$ adducts models}
In the following sections we present the theoretical models and simulated IR spectra of several PAH-C$_{60}$ adducts containing multiple PAH units; in many cases they were described by more than one spatial configuration (isomers). The mono-adducts notation refers to those C$_{60}$ adducts formed by the addition of one PAH unit only, while bis-adducts and tris-adducts stand for two and three PAH units, respectively. For the largest PAHs considered here, tetracene (Tn) and pentacene (Pn) (see Figure \ref{fig1}), only a maximum of two units and one unit, respectively, were attached to the C$_{60}$ cage; mainly due to our computational limitations.

\subsection*{PAH-C$_{60}$ adducts comparison}
\begin{figure}
    \centering
    \includegraphics[width = 8.5cm]{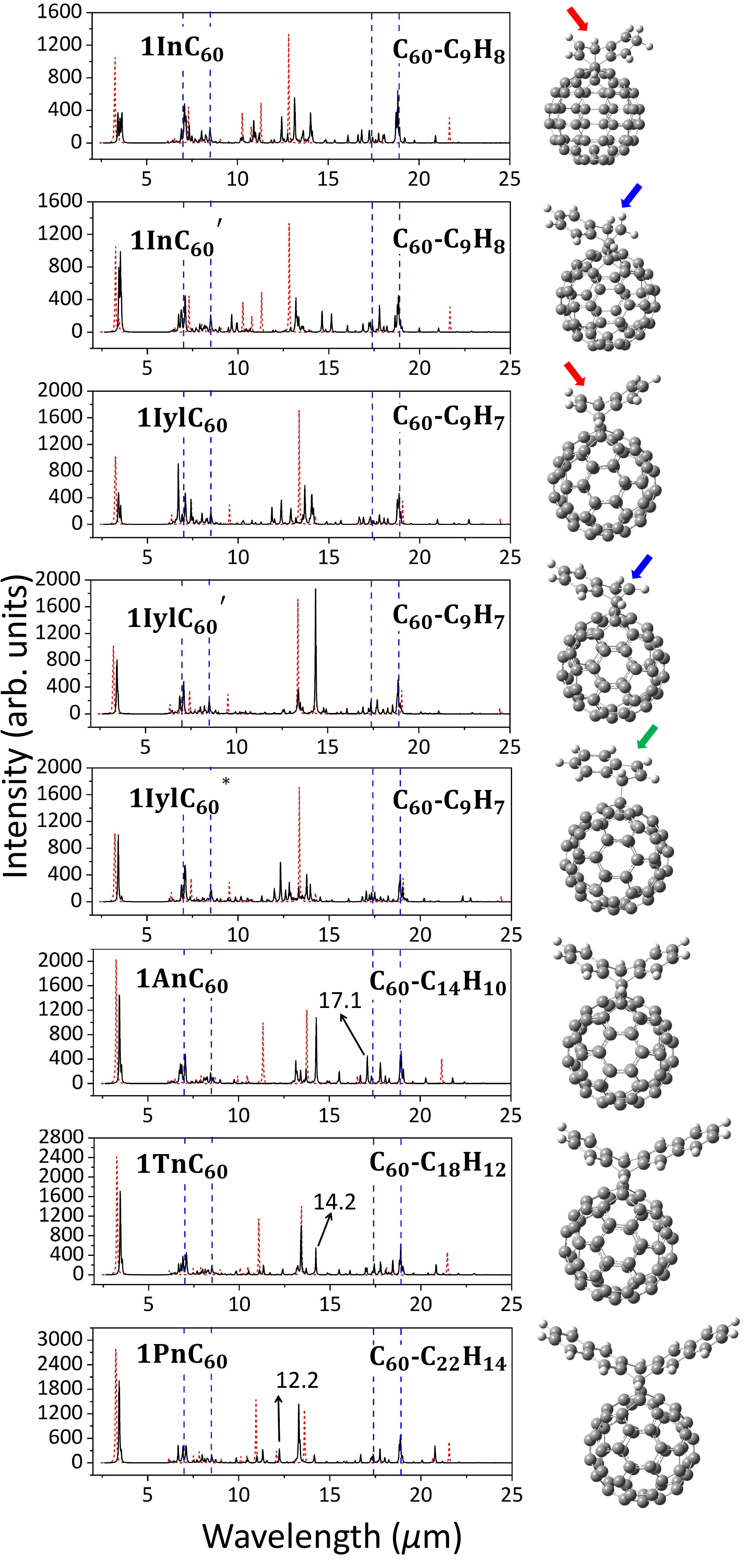}
    \caption{Scaled theoretical IR spectra of the mono PAH-C$_{60}$ adducts (one PAH unit: $\rm C_{60}$-$\rm C_x H_y$) modeled at the B3LYP-D3/6-31+G* level. All spectra were convolved with a Lorentzian function of FWHM = 0.02 $\mu$m. In the case of 2H-indene (In) and indenyl (Iyl), the blue and red arrows highlight the binding of the PAH with a pentagon and hexagon ring to the C$_{60}$ cage, respectively. For indenyl, with a single bond connecting its pentagon to C$_{60}$, a green arrow has been used instead. In all panels, the red dashed spectra correspond to the isolated pristine PAH, while the blue dashed lines mark the four strongest C$_{60}$ features ($\sim$7.0, 8.5, 17.4 and 18.9 $\mu$m). The black arrows on the spectra indicate those characteristic features (specific of each adduct), which are free of contribution from C$_{60}$, the pristine PAH and even from the other adducts.}
    \label{fig4}
\end{figure}

Figure \ref{fig4} displays the theoretical IR spectra of the different mono-adducts models together with their molecular structure. At first glance, the IR spectra of the mono-adducts are richer than those corresponding to the isolated PAH and C$_{60}$ molecules (Figure \ref{fig4}). Especially, in those spectral regions (e.g. the $\sim$12-13 and 14-17.5 $\mu$m regions, among others, see Figure \ref{fig4} and Appendix \ref{ap2}) where no contributions from the corresponding PAH and C$_{60}$ are observed. In addition, the typical $\sim$3.3 $\mu$m feature of the pristine PAH appears red-shifted in the case of PAH-C$_{60}$ adducts. The PAH binding to the C$_{60}$ cage is the cause for all new emission features observed in the IR spectra, denoting several unique spectral regions to distinguish PAH-C$_{60}$ mono-adducts. The black arrows in Figure \ref{fig4} also indicate those specific features that are free of contribution from C$_{60}$, the pristine PAH and even from the other adducts. In some adducts, like those with In and Iyl, it is very difficult to identify unique (specific of the adduct)  features since they contribute throughout all the IR spectra. The peculiarities of In or Iyl adducts are due to the multiple possible ways of binding to the C$_{60}$ cage. Single and double-bond bindings have also been found in previous studies about C$_{60}$-coronene$^+$ adducts, but in this case inducing the loss of one or two H atoms, which can be another types of binding between PAHs and C$_{60}$ reacting under energetic conditions \citep[see Figure 2 in][]{Dunk2013}. In contrast, the rest of mono-adducts could only be obtained under a double-bonded geometry (see Figure \ref{fig4}). Nevertheless, all mono-adducts have common emitting spectral regions with features at $\sim$3.3-3.6, 6-10, 12-16, and 17-19 $\mu$m.

The 3.3-3.6 $\mu$m feature can become broader (or splitted into various peaks) and less intense as the size of the PAH attached to the C$_{60}$ cage decreases (see Appendix \ref{ap2}). The former trend is only an indication of the reduction in the number of CH bonds and a breakdown of the symmetry as the PAH is smaller. Less CH bonds reduce the intensity of the 3.3-3.6 $\mu$m feature, while asymmetric PAHs like In and Iyl make the feature broader (or even with additional resolved features at slightly longer wavelengths of $\sim$3.5-3.6 $\mu$m) because of the different chemical environments affecting the CH bonds.\footnote{In the case of pure hexagon PAHs like anthracene (An), tetracene (Tn) and pentacene (Pn), the maximum different CH bonds in the mono-adducts is two, which are those CH close and far from the binding. In contrast, 2H-indene (In) and indenyl (Iyl) can have more than four different CH bonds due to the presence of pentagons and hexagons, broadening the 3.3-3.6 $ \mu$m feature. This is quite noticeable for In, which contains CH bonds connected to a C$\rm _{sp_3}$, creating multiple chemically different CH stretching.} In fact, In and Iyl show a clear increase in the IR intensity of the CH stretching region when the PAHs are bonded through the pentagon ring; as a consequence of a more marked dipole moment change for the CH vibrations inside the pentagon than for the hexagon.

At 6-10 $\mu$m the trend is similar in all spectra, a number of new low-intensity IR features; with  1InC$_{60}$$^{'}$ being the spectrally richer model (see Figure \ref{fig4}). The C-C vibrations lay in this region, which, as expected, shows the major coincidence (in terms of features) with its progenitors the free PAHs and the C$_{60}$ cage. In particular, all mono-adducts strongly contribute to the 7.0 $\mu$m C$_{60}$ feature, with almost a negligible emission contribution at 8.5 $\mu$m. The enrichment of the IR spectra in the 6-10 $\mu$m spectral region is smaller than for other fullerene based species like metallofullerenes \citep{Barzaga2023,Barzaga2023a}. This indicates that the dipole moment change related to C-C vibrations is weakly modified by the binding between the PAHs and C$_{60}$. Consequently, charge reordering and charge transfer processes in the mono-adducts are less important than for metallofullerenes \citep{Barzaga2023a}.

The 12-16 $\mu$m spectral region is very interesting because it is free from C$_{60}$ emission and the isolated PAHs usually only display one strong feature around $\sim$13-14 $\mu$m, but their mono-adducts counterparts display richer spectra. This region is mainly dominated by the out-of-plane CH vibrations, which is less IR active when the PAH is alone. In the particular case of 2H-indene (In) and indenyl (Iyl), the PAH can bind to the C$_{60}$ cage through its hexagon or pentagon ring\footnote{In the case of tetracene (Tn), we have also considered another possible way of binding, but it is highly thermodynamically unstable and also lacks of chemical principles (see Appendix \ref{ap1} for more details).} (see Figure \ref{fig4}). Both In and Iyl bind more favorably to the C$_{60}$ cage through the pentagon ring with energies of -1.62 eV (1InC$_{60}$$^{'}$) and -1.41 eV (1IylC$_{60}$$^{*}$), respectively. This specific binding is also reflected in the change of the IR spectra in the 12-16 $\mu$m region, with the appearance of new IR active features. On the other hand, in the case of Iyl single-bonded to the C$_{60}$ (1IylC$_{60}$$^{*}$, marked with a green arrow in Figure \ref{fig4}), multiple out-of-plane CH vibrations become IR active. This is due to the loss of symmetry compared to the doubled-bonded 1IlylC$_{60}$$^{'}$ model; the vibrational modes appear splitted, producing a richer spectra in the 12-16 $\mu$m range. 

Finally, another interesting spectral region is the one from 17 to 19 $\mu$m. This region looks quite similar in the mono-adducts, with several new weak features emerging around 17.4 $\mu$m and a broader 18.7-19.2 $\mu$m peak whose strongest intensity is at 18.9-19.0 $\mu$m. Clearly, this implies an important contribution to the 18.9 $\mu$m C$_{60}$ feature, with only a marginal emission contribution at 17.4 $\mu$m where there is a marked decay in intensity (see Appendix \ref{ap2}). The 18.9 $\mu$m feature of pristine C$_{60}$ corresponds to a combined carbon cage vibration, but the symmetry is destroyed by PAH binding.\footnote{This occurs for all the PAH-C$_{60}$ adducts under study, with the only exception of Iyl, which shows a broadness reaching up to 19.2 $\mu$m due to the combination with the intrinsic PAH vibrations.} Curiously, the largest mono-adducts from Tn and Pn are the only ones exhibiting noticeable IR features at wavelengths $>$20 $\mu$m; in particular, 1PnC$_{60}$ with a relatively strong feature at 20.7 $\mu$m. These features at long wavelengths are intrinsic to the free PAHs, although they are perturbed by binding to C$_{60}$, causing a shift in wavelength up to almost $\sim$0.8 $\mu$m in the case of Tn (see Figure \ref{fig4}).

\begin{figure}
    \centering
    \includegraphics[width = 7.5cm]{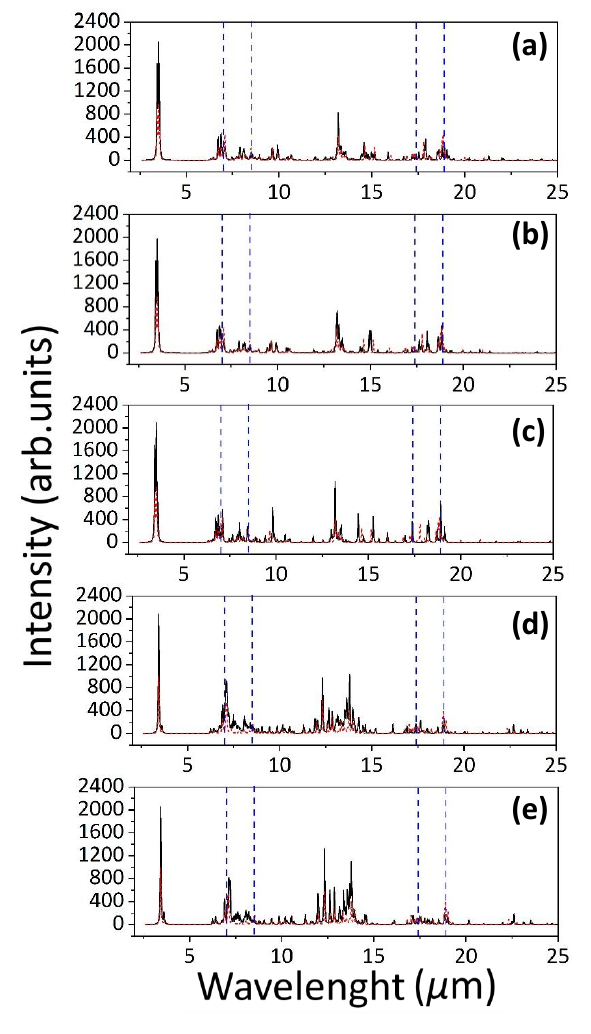}
    \caption{Theoretical scaled IR spectra of PAH-C$_{60}$ bis-adducts from 2H-indene (2InC$_{60}$, a-c) and indenyl (2IylC$_{60}$, d-e). The red dashed spectra correspond to the mono-adduct analogues, while the blue dashed lines mark the four strongest C$_{60}$ features ($\sim$7.0, 8.5, 17.4 and 18.9 $\mu$m). Note that the convolution parameters are the same as in Figure \ref{fig4}. For a detailed discussion on the models and spectral features we refer to Appendix \ref{bis}.}
    \label{fig5a}
\end{figure}

Figure \ref{fig5a} shows a comparison between the theoretical scaled IR spectra of mono-adducts and bis-adducts, indicating that no significant differences are observed. The changes from mono- to bis-adducts are only in terms of IR intensity caused by the increment of PAHs units, but regarding the features position, the spectra remain unchangeable. The only exception is for the model 2IylC$_{60}$ \textit{h} and is provoked by a strong structural deformation of the C$_{60}$ cage (see Appendix \ref{bis} for more details and also Appendix \ref{ap2}). Furthermore, the same spectral behavior is observed for the tris-adducts when they are compared with the bis-adducts (see Appendix \ref{tris}). The spectral differences among all these PAH-C$_{60}$ adducts are more appreciated at low IR intensity, which, in most cases, are expected to be well below the detection threshold of astronomical IR observations (see Appendices \ref{bis} and \ref{tris}).

\section*{PAH-C$_{60}$ adducts stability and kinetics}
\subsection*{Thermodynamic stability}\label{thermo}

The common mechanism for the formation of C$_{60}$ adducts is by exohedral addition of new molecular species, specially alkyl or benzyl groups, which is due to the high reactivity of C$_{60}$ \citep{hirsch1995,Krusic1991,li2022}. We can not confirm or refute that exohedral addition is the appropriate chemical process taking place in C$_{60}$-rich astrophysical environments like young PNe, characterized by both low temperatures \citep[$\sim$300 K; see e.g.][]{Cami2010,garcia2012} and densities; but it is the most likely one taking into consideration the well-known C$_{60}$ reactivity. In our case, multiple PAHs addends may undergo this exohedral cyclo-addition through sequential reaction pathways:
\begin{enumerate}[label=\Roman*., start=1]
    \item PAH+C$_{60} \xrightarrow[]{}$ PAH-C$_{60}$
    \vspace{0.05em}
    \item PAH+PAH-C$_{60} \xrightarrow[]{\rm +PAH}$ $\rm [PAH]_2$-C$_{60}$
    \vspace{0.05em}
    \item PAH+$ \rm [PAH]_2$-C$_{60} \xrightarrow[]{\rm +PAH}$ $\rm [PAH]_3$-C$_{60}$
    \vspace{0.5em}
\end{enumerate}
The three pathways (I-III) imply the sequential (step-by-step) addition of PAHs to C$_{60}$ from mono- to tris-adducts. These reactions depend on different factors that determine the mechanism of formation, but also lack the description of UV radiative processes, which are commonly observed in many astrophysical sources, including C$_{60}$-rich environments. Such analysis requires a detailed description of the several reaction pathways with the inclusion of photo-dissociation and photo-reaction, which involves the development of precise reactivity models; something that is extensively time-consuming and beyond of the scope of the present work. Thus, we have opted here for the derivation of the thermodynamic stability in order to identify the most spontaneous reactions following the sequential pathways (I-III). Assuming this approach implies that reactions take place under thermodynamical equilibrium. The former could be in contradiction with the observational data that indicate uncertainties in the estimation of physical conditions of astrophysical environments \citep[see e.g.][]{brieva16}, but our models can still be used as a rough prediction. Thus, we have chosen the difference in Gibb's free energy of formation per PAH unit ($\rm \Delta G_f$), defined by:
\begin{equation}
   \rm \Delta G_f = \frac{\sum G_{products}-\sum G_{reactants}}{n}
\end{equation}
where $\rm G_{products}$ and $\rm G_{reactants}$ are the Gibb's free energy of products and reactants, respectively, which have to be summed according to the reactions pathways I-III mentioned above. The variable n indicates the number of PAH molecules attached to the C$_{60}$ cage and is used to normalize the energy with respect to the PAHs units. A negative and positive value of $\rm \Delta G_f$ indicates a spontaneous and non-spontaneous process, respectively. It is important to clarify that $\rm \Delta G_f$ is a thermodynamic quantity that describes the probability that reactants become products; however, there is no information on the activation energy to carry out these reactions. In this sense, even though $\rm \Delta G_f$ provides insight into the binding energy of the PAH on C$_{60}$, it can not be used as a threshold for any reaction mechanism (e.g. dissociation, photodissociation, etc.). Theoretical studies on the C$_{60}$-anthracene$^+$ adduct suggest that the energy barriers in PAH binding to C$_{60}$ can be as small as 0.08 eV, which implies that the activation energy is almost equal to the binding energy \citep[see Figure 6 in][]{Zhen2019a}. In principle, this could be used to infer the range in activation energies of our PAH-C$_{60}$ adducts because they also follow a Diels-Alder binding. However, it is likely that differences in the thermodynamics could appear due to the fact that our models are neutral systems. Table \ref{tb2} summarizes the thermodynamic data for all PAH-C$_{60}$ adducts under study; all values have been computed assuming a temperature of 300 K and adequately corrected by the zero-point energy. 

\begin{table}
\centering
      \caption[]{Thermodynamic values of all the PAH-C$_{60}$ adducts}
         \label{tb2}
         \setlength\tabcolsep{0.8pt}
         \renewcommand{\arraystretch}{1.1}
         \begin{tabular}{cccccccc}
            \hline
            \noalign{\smallskip}
             &  \multicolumn{3}{c}{Mono-adduct} & \multicolumn{2}{c}{Bis-adduct} & \multicolumn{2}{c}{Tris-adduct}\\
            \noalign{\smallskip}
            PAH &  Model & $\rm E_b$ & $\rm\Delta G_f (I)$ & Model & $ \rm \Delta G_f (II)$ & Model & $\rm \Delta G_f (III)$\\
            \noalign{\smallskip}
            \hline
            \noalign{\smallskip}
            \multirow{3}{*}{In} & 1InC$_{60}$$^{'}$ &-0.11 & -1.015 & 2InC$\rm _{60}$ {\it a} &-0.976 & 3InC$\rm_{60}$ {\it a} & -0.953\\
        
                        &  &   &  &  2InC$\rm_{60}$ {\it b} & --1.010 &  & \\
        
                        &  &   &  &  2InC$\rm_{60}$ {\it c} & -0.978& 3InC$\rm_{60}$ {\it b} & -0.974 \\\arrayrulecolor{gray} \cline{2-8}
             \noalign{\smallskip}
             \multirow{4}{*}{Iyl} & 1IylC$_{60}$$^{'}$ &0.35& 1.000 & 2IylC$\rm_{60}$ {\it g} & 1.820 & 3IylC$\rm_{60}$ {\it c} & 1.052\\
              
              && &  &2IylC$\rm_{60}$ {\it f} & 1.022 &3IylC$\rm_{60}$ {\it d}  & 1.075\\
              
              & 1IylC$_{60}$$^{*}$ &-0.46& 0.102 &  2IylC$\rm_{60}$ {\it e} & -0.056 &  & \\
            
              &&  &  &  2IylC$\rm_{60}$ {\it d} & 0.075 &3IylC$\rm_{60}$ {\it e} & -0.064  \\\cline{2-8}
              \noalign{\smallskip}
             \multirow{2}{*}{An} & 1AnC$_{60}$ &-0.49& 0.121 & 2AnC$\rm_{60}$ {\it h} & 0.124 & 3AnC$\rm_{60}$ {\it f} & 0.183 \\
                
                     &  &    &  &  &  & 3AnC$\rm_{60}$ {\it g} & 0.154 \\\cline{2-8}
              \noalign{\smallskip}
                \text{Tn} &  1TnC$_{60}$ &-0.33& -3.261 & 2TnC$\rm_{60}$ {\it i} &-2.389 &  \\\cline{2-8}\noalign{\smallskip}
                \text{Pn} &  1PnC$_{60}$ &-1.05& -0.426  \\
            \arrayrulecolor{black}\hline
         \end{tabular}
         \tablefoot{Gibb's free energy of formation per PAH unit ($\rm\Delta G_f$) for each of the PAH-C$_{60}$ adducts studied here (Figures \ref{fig4}, \ref{figb1} and \ref{figb2}). The binding energies ($\rm E_b$) for the mono-adducts are also indicated. Each $\rm\Delta G_f$ value corresponds to the sequential reaction pathways (I-III). All energies are described in eV units.}
         \end{table}
According to the values in Table \ref{tb2}, tetracene (Tn) exhibits the most spontaneous (exergonic) reactions for any of the reaction pathways assumed. Increasing the number of Tn addends makes the reactions less exergonic. Such trend in $\rm \Delta G_f$ is seen for almost all PAHs, with the exception of some C$_{60}$ adducts with In and Iyl. This generally indicates that the inclusion of more PAH units to C$_{60}$ is energetically hindered by the geometrical arrangement of multiple PAHs to the carbon cage, which is known as steric effect. In contrast, the thermodynamics of the indenyl (Iyl) adducts is the most complex one since there is no clear relationship between the $\rm \Delta G_f$ values and the reaction pathways and/or regioisomer (Table \ref{tb2}). However, it is clear that the single-bonded Iyl-C$_{60}$ adducts models are more exergonic than the double-bonded ones (see all adducts models from 1IylC$_{60}$$^{*}$ in Table \ref{tb2}). Indenyl is a planar and highly unsaturated radical that tends to change from C$_{\rm sp_2}$ to C$_{\rm sp_3}$ when it binds to C$_{60}$. A similar behavior is known to occur for other aromatic addends \citep[e.g.][]{Mas-Torrent2002}. 

Our predictions also show that the formation of full hexagon PAH-C$_{60}$ adducts with anthracene (An) is an exothermic process (see $\rm \mathbf{E_b}$ in Table \ref{tb2}); similar theoretical results have been reported for C$_{60}$-anthracene$^+$ derivatives in recent studies \citep{Zhen2019,Wu2024}. Steric and electrostatic effects should be the main factors for the thermodynamic changes when the amount of PAH addends increases, but our models contain both symmetric and asymmetric PAHs, and the thermodynamics information is thus not straightforward to interpret. Seemingly, there is no correlation between the increment in the size of fused-hexagon PAHs (An $<$ Tn $<$ Pn) and the observed thermodynamics (see Table \ref{tb2}). Future intensive efforts would be necessary to include several factors in the mechanistic description of these chemical reaction pathways.

\subsection*{Kinetics of PAH-C$_{60}$ adducts}\label{kinetic}
Although in the present study no precise reactivity models have been developed to predict the kinetics of our PAH-C$_{60}$ adducts, previous experimental works have determined these parameters for some of the adducts presented here. This Section covers the state-of-the-art experimental kinetics, concerning the rate constants of the mono- and bis-adducts of C$_{60}$ with anthracene, tetracene and indene.

Kinetics analyses of PAH-C$_{60}$ adducts formation have been mainly made during Diels-Alder synthesis, which is the common experimental method to obtain them.  \citet{SAROVA2004} study the chemical kinetics of C$_{60}$ mono-adducts from anthracene and tetracene combining experimental measurements and theoretical predictions. By using a fitting procedure, they determine rate constants following a second-order kinetics with values of 1.6$\times$10$^{-4}$M$^{-1}$s$^{-1}$ and 3.1$\times$10$^{-2}$M$^{-1}$s$^{-1}$ for anthracene and tetracene, respectively, at T$\sim$298 K \citep{SAROVA2004}. Such values indicate that the tetracene mono-adduct kinetics is much faster than for anthracene at least by a factor of 200. As we will see in the next Section, these kinetics data can be used, at least when the small PAHs like anthracene and tetracene are equally abundant, for the construction of more reliable theoretical IR spectra representative of PAH-C$_{60}$ adducts.

\begin{figure}
    \centering
    \includegraphics[width = 8.7cm]{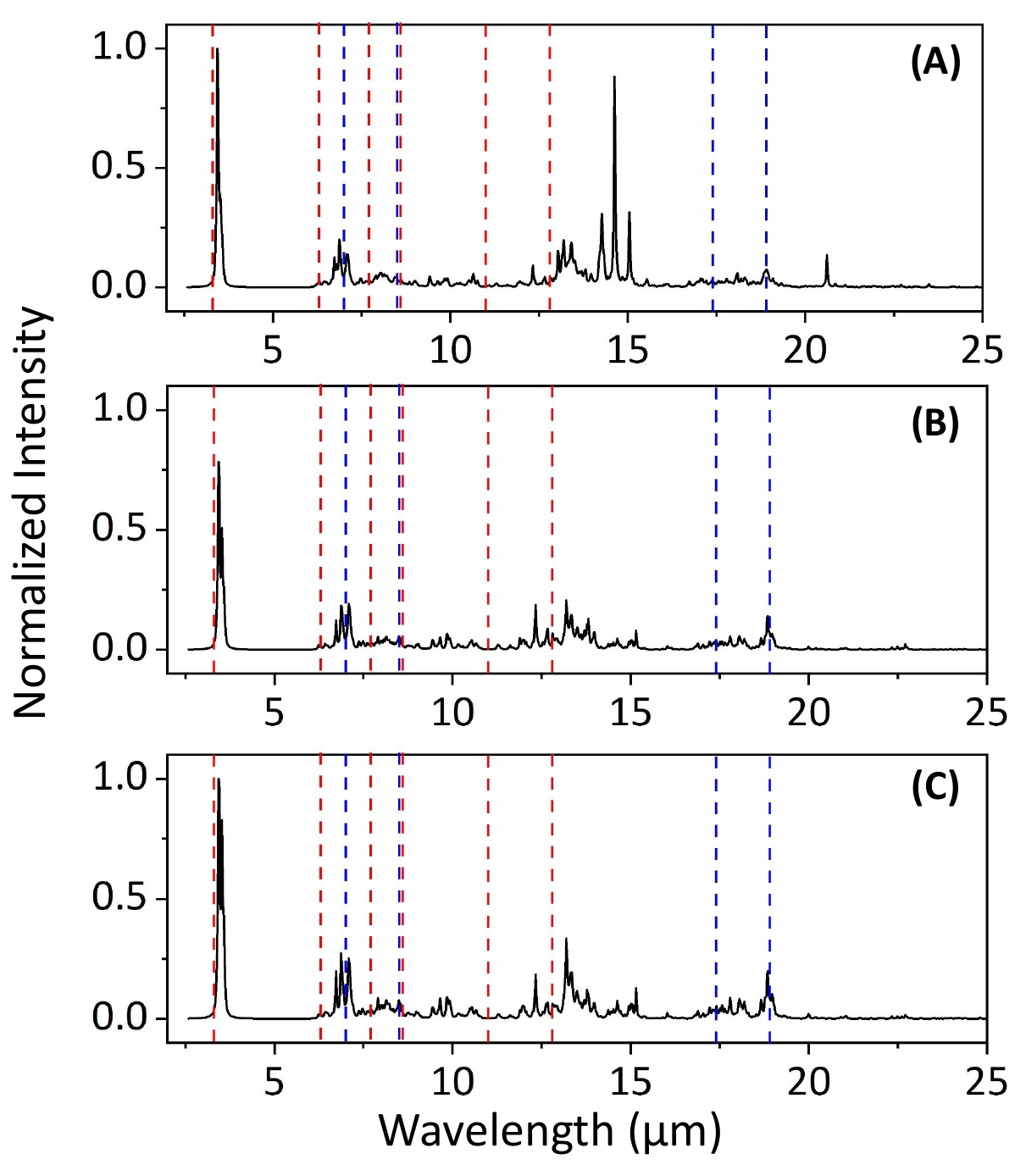}
    \caption{The DFT simulated IR ($\sim$3-25 $\mu$m) spectra of the mixture of PAH-C$_{60}$ adducts. (A): the total (summed) mixture spectrum; (B): the total abundance-weighted spectrum using the indene abundance from \citet{Burkhardt2021}; and (C) the total abundance-weighted spectrum using the indene abundance from \citet{Cernicharo2021}. In all panels, the red dashed lines mark the most common astronomical PAH features (at $\sim$3.3, 6.2, 7.7, 8.6, 11.2 and 12.7 $\mu$m), instead blue dashed lines mark the four strongest C$_{60}$ features ($\sim$7.0, 8.5, 17.4 and 18.9 $\mu$m). In the case of total the (summed) mixture spectrum, the intensity has been normalized to the maximum intensity peak, while the abundance-weighted spectra have been normalized with respect to the higher intensity peak from the weighted spectrum (C); i.e. the one using the indene abundance from \citet{Cernicharo2021}.}
    \label{fig8}
\end{figure}

More recently, the kinetics of solid indene-C$_{60}$ mono- and bis-adducts decomposition have been reported by \citet{RODRIGUES2023}. The most remarkable result of the previous study is the relative rate constants between the mono- and bis-adducts, being one of the few scarce data of this type reported in the literature. Accordingly, a first-order kinetics defines the decomposition of mono-adducts carrying out a slower process with respect to bis-adducts (at least by a factor of 2). Arguably, solid-state kinetics can differ from the gas-phase since different forces may rule out the reaction. We are thus aware that the kinetics data by \citet{RODRIGUES2023} could be far from the chemical processes occurring in astrophysical environments. However, they can be still used to build more reliable and chemical intuitive preliminary models, describing the IR emission of these PAH-C$_{60}$ adducts and elaborated below.

\section*{Astrophysical relevance}\label{AR}
A comparison between our theoretical spectra of PAH-C$_{60}$ adducts and the observational ones is not straightforward because it is well known that the relative abundances of the several species may play a key role in the landscape of the IR spectra \citep[see e.g.][]{Barzaga2023,Barzaga2023a}. Also, all previous IR spectroscopic observations of fullerene-rich astrophysical environments were carried out by the {\it Spitzer Space Telescope}, which did not cover the $\sim$3-5 $\mu$m region. As we will see through the current Section, this is a key spectral region where the PAH-C$_{60}$ adducts strongly contribute to the IR emission.

The main possible drawback of the kinetics information in the previous Section could be that the rate constants are obtained from experimental reactions where anthracene and tetracene always exceeded C$_{60}$. The abundances of such small PAHs and C$_{60}$ in astrophysical environments are not precisely known; mainly because their more specific UV/Visible electronic transitions have not been detected yet and only upper limit estimates to their abundances\footnote{Note that the C$_{60}$ abundance estimates from the IR emission usually assume thermal emission as excitation mechanism \citep[see e.g.][and references therein]{Cami2010,garcia2012} and the IR bands attributed to C$_{60}$ may actually comprise contributions by analogous species like metallofullerenes \citep[see e.g.][]{Barzaga2023,Barzaga2023a}. In fact, \citet{brieva16} concluded that thermal excitation by itself cannot explain the IR spectra attributed to C$_{60}$ in some planetary nebulae, and also that a possible explanation is that other molecules than C$_{60}$ may contribute to the observed spectra.} are available \citep[see][for a recent review about C$_{60}$]{rouille2021}. Another difficulty is that present abundance estimates are not coming from the same astrophysical source (e.g. the diffuse ISM, PNe, RCBs, etc.) and/or astronomical spectra. By searching in the literature, we found diffuse ISM abundance-limit estimates of both anthracene and C$_{60}$ towards the same sources (the reddened stars HD 169454 and HD 183143) and using the same astronomical spectra \citep[][]{gredel2011,rouille2021}. The upper limits for column densities of interstellar C$_{60}$ and anthracene are in the ranges $\sim$1$-$8 $\times$ 10$^{12}$ cm$^{-2}$ and 1$-$3 $\times$ 10$^{12}$ cm$^{-2}$, respectively. Other small PAHs like pyrene and 2,3-benzofluorene studied by \citet{gredel2011} display upper limits for column densities on the same order ($\sim$2$-$8 $\times$ 10$^{12}$ cm$^{-2}$). The upper limits for C$_{60}$ column densities in the ISM are very similar to those estimated towards the C$_{60}$-rich circumstellar environments around PNe and RCB stars \citep[$\sim$1$-$4 $\times$ 10$^{12}$ cm$^{-2}$,][]{garcia2012b,garcia2013,diaz2015}. In short, with the UV/Visible astronomical data at hand, it is not possible to know the relative abundances of C$_{60}$ and small PAHs in astrophysical environments like the diffuse ISM and C$_{60}$-rich circumstellar envelopes. However, recent radioastronomy observations have detected indene (In) towards the cold dark cloud TMC-1 \citep[e.g.][]{Cernicharo2021}, with a surprisingly high column density of $\sim$1.6 $\times$ 10$^{13}$ cm$^{-2}$ \citep[a slightly lower value of $\sim$0.96 $\times$ 10$^{13}$ cm$^{-2}$ was obtained by][]{Burkhardt2021}. Although it is not known if TMC-1 harbor both C$_{60}$ and indene, such observations suggest that some small PAHs can be found in very high abundance (even higher than C$_{60}$) in space, as long as they are shielded from strong (inter)stellar UV photons \citep[e.g.][]{Stockett2025}. We note that such a possible UV-shielding, permitting small PAHs to survive, may not be the case for all interstellar and circumstellar environments where fullerenes (C$_{60}$, C$_{60}$$^{+}$ and C$_{70}$) have been detected. However, these shielding conditions (e.g. UV-shielding by circumstellar dust grains) are usually met in evolved stars (sources between AGB stars and PNe) and neutral molecules, both simple (H$_{2}$, CO) and complex (PAHs, C$_{60}$) ones, may be formed in the outer parts or within clumps \citep[see e.g.][]{manchado2015,wesson2024,gold2024,Clark2025}; indeed neutral C$_{60}$ seems to be distributed in rings and/or clumps in young PNe \citep{diaz-luis2018,cami2018}. In addition, for the particular case of the circumstellar environments of evolved stars, there is a rapidly changing UV radiation field (from no or very little UV photons in the AGB phase to ionizing UV photons in the PNe phase). At present, it is not known at which exact AGB-PNe evolutionary stage PAHs and fullerenes are formed. It is only known that fullerenes are detected for a very short time when the UV photons from the central star are strong enough to photoionize H (central stars with T$_{eff}$$\sim$30$-$40 kK) but the excitation mechanism (e.g. fluorescence vs thermal) of the C$_{60}$ IR emission is unknown \citep[e.g.][]{brieva16}. In other words, fullerenes could be formed at shielding conditions (and thus coexist with small PAHs for a certain time) but they only could be detected when excited by relatively strong UV photons and emit in the mid-IR; although this is not always the case and C$_{60}$ emission is also detected in astrophysical environments with rather little UV radiation like proto-PNe and RCB stars. 

Therefore, we tentatively estimate the possible order of magnitude of the expected abundances of PAH-C$_{60}$ adducts by using the thermodynamics\footnote{In this case, we have used the enthalpy energy, which is directly obtained from our quantum-chemistry calculations as well as the free Gibb's energies listed in Table \ref{tb2}.} and kinetics data presented in the previous Sections (see the Appendix \ref{ap} for more details). We note that our abundance estimation assumes the average thermal conditions of the C$_{60}$-rich circumstellar envelopes around evolved stars (T$\sim$300 K) but uses the estimated column densities of indene towards TMC-1 \citep[as obtained from extremely sensitive radio observations,][]{Burkhardt2021,Cernicharo2021}. This means that our abundance estimates may be useful as rough illustrations only, because the physical conditions in the dense molecular cloud TMC-1 can be very different to those in C$_{60}$-rich circumstellar environments. This is due to the lack of indene abundance estimates toward fullerene-rich circumstellar environments; e.g. similar extremely sensitive radio astronomical observations towards C$_{60}$-rich environments are not available in the literature. The DFT simulated mixture spectra in Figure \ref{fig8} have been constructed following a procedure similar to that in \citet{Barzaga2023,Barzaga2023a}; see Appendix \ref{ap}\footnote{In the case of the total abundance-weighted spectra, they both have been normalized with respect to the maximum intensity peak obtained from the \citet{Cernicharo2021} indene abundance data (spectrum C in Figure \ref{fig8}). This way, it is noticeable the change in IR intensity with respect to the lower indene abundance from \citet{Burkhardt2021} (spectrum B in Figure \ref{fig8}).} for more details. 

From Figure \ref{fig8}, it is very clear the important effect of considering the PAH-C$_{60}$ abundances in the resulting IR spectra. The pure summed PAH-C$_{60}$ mixture in Figure \ref{fig8}a exhibits a spectrum, mainly characterized by strong emission features at $\sim$3.4-3.6, 14.2, 14.7 and 15.1 $\mu$m. These features are characteristic of the Iyl bis-adduct under configuration {\it g} (Figure \ref{figb1}g), which is highly unstable. This landscape drastically changes when our estimated PAH-C$_{60}$ abundances are considered in the simulations. In particular, there is a strong modification in the 3.4-3.6 $\mu$m feature, which becomes splitted in the abundance-weighted PAH-C$_{60}$ mixtures, with distinctive peaks at 3.43, 3.51 and 3.57 $\mu$m (see Figures \ref{fig8}b, c and Appendix \ref{ap2}). Furthermore, the abundance-weighted spectra are spectrally richer along the full $\sim$5-20 $\mu$m spectral range in terms of distinguishable peaks, showing numerous features weaker than the 3.4-3.6 $\mu$m one, but still noticeable. The most intense IR features in the latter spectral range are those at $\sim$6.7-7.4, 12.2-14.0 and 18.6-19.2 $\mu$m. In short, the strong differences seen in the DFT spectra in Figure \ref{fig8} highlight the importance of considering, at least, the tentative abundances of the several species in the theoretical simulation of IR spectra from PAH-C$_{60}$ adducts.

Interestingly, there is a common factor in the spectra of Figure \ref{fig8}, which is the general absence of IR contribution at the wavelengths of the astronomical PAHs (at $\sim$6.2, 7.7, 8.6, 11.2, and 12.7 $\mu$m) and the C$_{60}$ features at $\sim$8.5 and 17.4 $\mu$m. Although both species are the progenitors of PAH-C$_{60}$ adducts, the resulting theoretical spectra with (by far) the most intense IR bands at $\sim$3.4-3.6 $\mu$m don't indicate their presence. From the astronomical point of view, this suggests that PAH-C$_{60}$ adducts could contribute to the $\sim$3.4-3.6 $\mu$m IR emission and usually attributed to aliphatic carbon species (see below). The astronomical PAH and C$_{60}$ features are well characterized and the astrophysical community has generally accepted that their detection indicates the presence of these species. However, according to our tentative predictions, the absence of IR emission from PAHs and/or C$_{60}$ does not necessarily mean that they are not present; they could be present, instead, forming more complex hybrid species such as PAH-C$_{60}$ adducts. A recent theoretical work by \citet{Xu2023} finds that much simpler C$_{60}$ adducts, like C$_{60}$H$^{+}$ and C$_{60}$O$^{+}$, exhibit most of the pristine C$_{60}$ features in their predicted IR spectra. Apparently, such a difference is due to the large diversity in the interactions between the PAHs and C$_{60}$. Thus, the predictions presented here suggest that one of the possible explanations for the lack of PAH and C$_{60}$ features in astronomical environments could be the formation of their adducts. 

Our theoretical calculations may also give some insights about the possible origin of the $\sim$3.4-3.6 $\mu$m IR features, sometimes observed in astronomical sources \citep[see e.g.][for a recent review]{li2020}; e.g. in proto-PNe, where satellite features from 3.4 to 3.6 $\mu$m, accompanying the 3.3 $\mu$m feature, have been observed \citep{geballe1992,Hrivnak2007,Materese2017}. Such 3-5 $\mu$m observations, however, have been carried out with space- and ground-based telescopes less sensitive than Spitzer, which otherwise didn't cover this spectral range.
 In particular, \citet{geballe1992} analyzed several objects in the short transition phase from AGB stars to PNe (i.e. proto-PNe) and observed remarkably strong 3.4-3.5 $\mu m$ emission relative to the usually dominant 3.3 $\mu m$ feature. They concluded that the observed 3.4-3.5 $\mu m$ features cannot be explained by the presence of stretch overtones of CH produced by PAHs and they suggested that they should be assigned to the stretching modes of aliphatic CH$_{2}$ and CH$_{3}$ groups. On the other hand, \citet{Materese2017} suggested that the correlation between the abnormally intense 3.4-3.5 $\mu m$ and 6.9 $\mu m$ features observed in this kind of post-AGB objects could be due to the presence of superhydrogenated PAHs (H$_n$-PAHs), containing aliphatic CH$_2$ and CH$_3$ moieties. The carbon chemistry in space is dominated by the 3.3 and 3.4 $\mu$m features or aromatic-aliphatic dichotomy, with aromatics (PAHs; 3.3 $\mu$m) generally believed to be a main component of carbon in space. This popular perception, however, could be challenged in the JWST era. Thanks to the exceptional sensitivity and spectral resolution of recent JWST observations, the 3.4 $\mu$m features are routinely detected in very different types of astronomical sources; many of them with a no significant UV radiation field like proto-PNe, trojans, active galactic nuclei, etc. \citep[see e.g.][]{Lai2023,wong2024} but also in more evolved and old PNe with a much stronger UV radiation field \citep{Clark2025}.
 
 Considerable efforts have been previously made to understand the origin (aliphatics, PAHs, etc.) of these unusual $\sim$3.4-3.6 $\mu$m features. The PAH molecules can barely fit these features by using a very large family of irregular structures, even assuming the effects of anharmonic couplings due to high temperatures \citep[see e.g.][]{joblin1995,Candian2012,Bauschlicher2009}. A more recent theoretical work suggests that IR features beyond 3.4 $\mu$m, which is the archetypal aliphatic CH stretching mode, cannot be assigned to pure CH bonds \citep{Sadjadi2017}. According to \citet{Sadjadi2017} those features should probably involve another element (e.g. N, S, O) besides C and H, contaminating the CH stretching. On the other hand, hydrogenated fullerenes (fulleranes) have also been suggested to contribute in the $\sim$3.4-3.6 $\mu$m region via their CH bonds \citep{Iglesias-Groth2012,Zhang2017}. This is reinforced by recent abundance predictions, suggesting that C$_{60}$H$^+$ should be abundant in the diffuse ISM \citep{Abbink2024}. Also, accurate experimental studies on fulleranes \citep[$\rm C_{60}H^+$, $\rm C_{70}H^+$;][]{palotas2020,Finazzi2024} and oxidized fullerenes \citep[$\rm C_{60}O^+$, $\rm C_{60}OH^+$;][]{Palotas2024} demonstrate that these species have a common IR emission region around $\sim$6.2-8.3 $\mu$m, corresponding to CC vibrations. Thus, since all these fullerene compounds contain CH and CC stretching, their IR spectra could be entangled with those from PAH-C$_{60}$ adducts in several spectral regions. However, a possible way to tackle this, in order to resolve or distinguish their contribution in astronomical IR spectra, is via the CH/CC stretching band ratio. For example, \citet{Finazzi2024} find that the CH/CC band ratio for fulleranes is quite small due to their low IR intensity of CH stretching; note that CH stretching is not present for oxidized fullerenes \citep{Palotas2024}. In contrast, the PAH-$\rm C_{60}$ adducts studied here display a CH/CC band ratio larger than fulleranes (see Figure \ref{fig3}). The potential use of the CH/CC band ratio to distinguish between different H-containing fullerene derivatives would require sensitive astronomical data like those from the JWST; i.e. covering the entire IR range ($\sim$3-25 $\mu$m) of molecular vibrations.

Clearly, there is no consensus yet about the carrier/s of the $\sim$3.4-3.6 $\mu$m features (and similarly occurs for the 6.9 $\mu$m feature generally observed in conjunction with them). Our experimental-theoretical study of PAH-C$_{60}$ adducts suggests other species that may contribute at these wavelengths (see Figures \ref{fig3}, \ref{fig8} and \ref{app4}). Remarkably, the PAH-C$_{60}$ adducts display strong $\sim$3.4-3.6 $\mu$m features without the presence of aliphatic CH bonds (with the exception of the indene derivatives), due to the interplay between C$_{60}$ and small PAHs (see Figures in the Appendix \ref{ap2}). In contrast to \citet{Sadjadi2017} and \citet{Zhang2017}, which use a diverse set of model structures, our set of PAH-C$_{60}$ adducts are rather structurally simple. It is thus likely that the contamination of PAH-C$_{60}$ adducts by aliphatic chains within the PAH structure would increase, even more, the intensity of the $\sim$3.4-3.6 $\mu$m IR features. Future efforts should enlarge the PAHs diversity, from linear to branched or even helical structures. 

\section*{Summary}
In summary, we have computed IR simulated spectra of multiple PAH-C$_{60}$ adducts based on experimental data combined with quantum-chemical calculations. According to the experimental data and theoretical results the formation of these new species would almost erase the presence of most of the characteristic IR features from the pristine PAHs and some from isolated C$_{60}$. Furthermore, we have very roughly estimated the possible abundances of PAH-C$_{60}$ adducts in C$_{60}$-rich astrophysical environments by applying the state-of-the-art kinetics data, hopefully building more reliable and global (or weighted) simulated IR spectra.  Consequently, the abundance-weighted spectra display a series of new relevant IR features like a broad $\sim$3.4-3.6 $\mu$m feature, a richer 6-10 µm spectral region, a strong modification within the 12-16 $\mu$m range together with emission contribution to the 7.0 and 18.9 $\mu$m C$_{60}$ features. Surprisingly, the PAH-C$_{60}$ adducts presented here, with almost no CH aliphatic in their structure, display strong $\sim$3.4-3.6 $\mu$m features and could be potential carriers of this kind of emission in astronomical environments; especially in those where C$_{60}$ is known to be abundant. Unfortunately, previous IR observations of C$_{60}$-rich astrophysical environments were done by Spitzer with no access to the 3-5 $\mu$m spectral range. However, such observations could be made at high-sensitivity by the JWST; with its spectral coverage well below 5 $\mu$m. This along with high-quality experimental IR spectra of diverse PAH-C$_{60}$ adducts (e.g., including those formed by C$_{60}$ and PAHs of different size and chemical nature) will give more insights on the possible IR emission contribution of PAH-C$_{60}$ adducts to the 3-4 $\mu$m spectral region.

\section*{Data availability}
All the scaled harmonic DFT spectra presented herein, as well as the IR cross sections used to construct them, are available at \url{https://doi.org/10.5281/zenodo.17485506}. 
\begin{acknowledgements}
      We acknowledge the support from the State Research Agency (AEI) of the Ministry of Science, Innovation and Universities (MICIU) of the Government of Spain, and the European Regional Development Fund (ERDF), under grants PID2020-115758GB-I00/AEI/10.13039/501100011033, PID2022-136970NB-I00/AEI/10.13039/501100011033 and PID2023-147325NB-I00/AEI/10.13039/501100011033. The authors also express their gratitude to the Deanship of Graduate Studies and Scientific Research at Taif University for their financial support of this work. This publication is based upon work from COST Action CA21126 - Carbon molecular nanostructures in space (NanoSpace), supported by COST (European Cooperation in Science and Technology). R.B and B.K also acknowledge the generous allocation of computer time at LaPalma-IAC Supercomputer and CITIC Universidade da Coruña.      
      This article used flash storage and CPU/GPU computing resources as Indefeasible Computer Rights (ICRs) being commissioned at the ASTRO POC project that Light Bridges will operate in the Island of Tenerife, Canary Islands (Spain). The ICRs used for this research were provided by Light Bridges in cooperation with Hewlett Packard Enterprise (HPE) and VAST DATA. The authors wish to acknowledge the contribution of the IAC High-Performance Computing support team and hardware facilities to the results of this research. BK is grateful to Science by Women program for funding a six-months visiting senior research fellowship to IAC. The authors acknowledge Deanship of Scientific Research, Taif University for funding this work.
\end{acknowledgements}

   \bibpunct{(}{)}{;}{a}{}{,}
   \bibliographystyle{aa} 
   \bibliography{aa54031-25} 

\begin{appendix}
\section{Empirical-corrected scaling factors}\label{SFemp}
The scaling factors reported in Table \ref{tb1} were empirically corrected in the $\sim$3.3 $\mu$m region to improve the accuracy with respect to the experimental spectra (see the corresponding Section and related discussion); Table \ref{tba1} shows the final values.
\begin{table}[h!]
      \caption[]{Frequency-range-specific scaling factors for the PAHs in Figure \ref{fig1}.}
         \label{tba1}
\centering        
         \begin{tabular}{lll}
            \hline
            \noalign{\smallskip}
            PAH & $\mathrm{SF_3^{D3}}$ &$\mathrm{SF_{3\,{\mathrm{emp}}}^{D3}}$\\
            \noalign{\smallskip}
            \hline
            \noalign{\smallskip}
                        & 0.9852$^{(1)}$ & 0.9852 \\
             \text{In} &  0.9675$^{(2)}$ & 0.9675 \\
                        & 0.9614$^{(3)}$ & \underline{0.9090} \\ \arrayrulecolor{gray} \noalign{\smallskip}\cline{2-3}
                        \noalign{\smallskip}
                        & 0.9838   & 0.9838 \\
             \text{Iyl} &  0.9664  &  0.9664\\
                        & 0.9627   & \underline{0.9103}  \\ \noalign{\smallskip}\cline{2-3}\noalign{\smallskip}
              
                       & 0.9807  & 0.9807  \\
             \text{An} & 0.9711 & 0.9711 \\
                       & 0.9610   & \underline{0.9086} \\ \noalign{\smallskip}\cline{2-3}\noalign{\smallskip}
              
                       &  0.9818 & 0.9818 \\
             \text{Tn} &  0.9679 & 0.9679 \\
                       &  0.9614  & \underline{0.9090} \\ \noalign{\smallskip}\cline{2-3}\noalign{\smallskip}
              
                       & 0.9848 & 0.9848 \\
             \text{Pn} & 0.9648 & 0.9648   \\
                       & 0.9618 & \underline{0.9094}   \\
            \noalign{\smallskip}
            \arrayrulecolor{black}\hline
         \end{tabular}
      \tablefoot{The scaling factors from our 6-31+G(d)/B3LYP+GD3 (SF$_3^{\mathrm{D3}}$) method and their corresponding empirical-corrected ($\mathrm{SF_{3\,{\mathrm{emp}}}^{D3}}$) are reported. The legends remain as in Table \ref{tb1}.}
   \end{table}

\FloatBarrier\section{Bis-adducts}\label{bis}
\begin{figure*}
    \centering
    \includegraphics[width = 17cm]{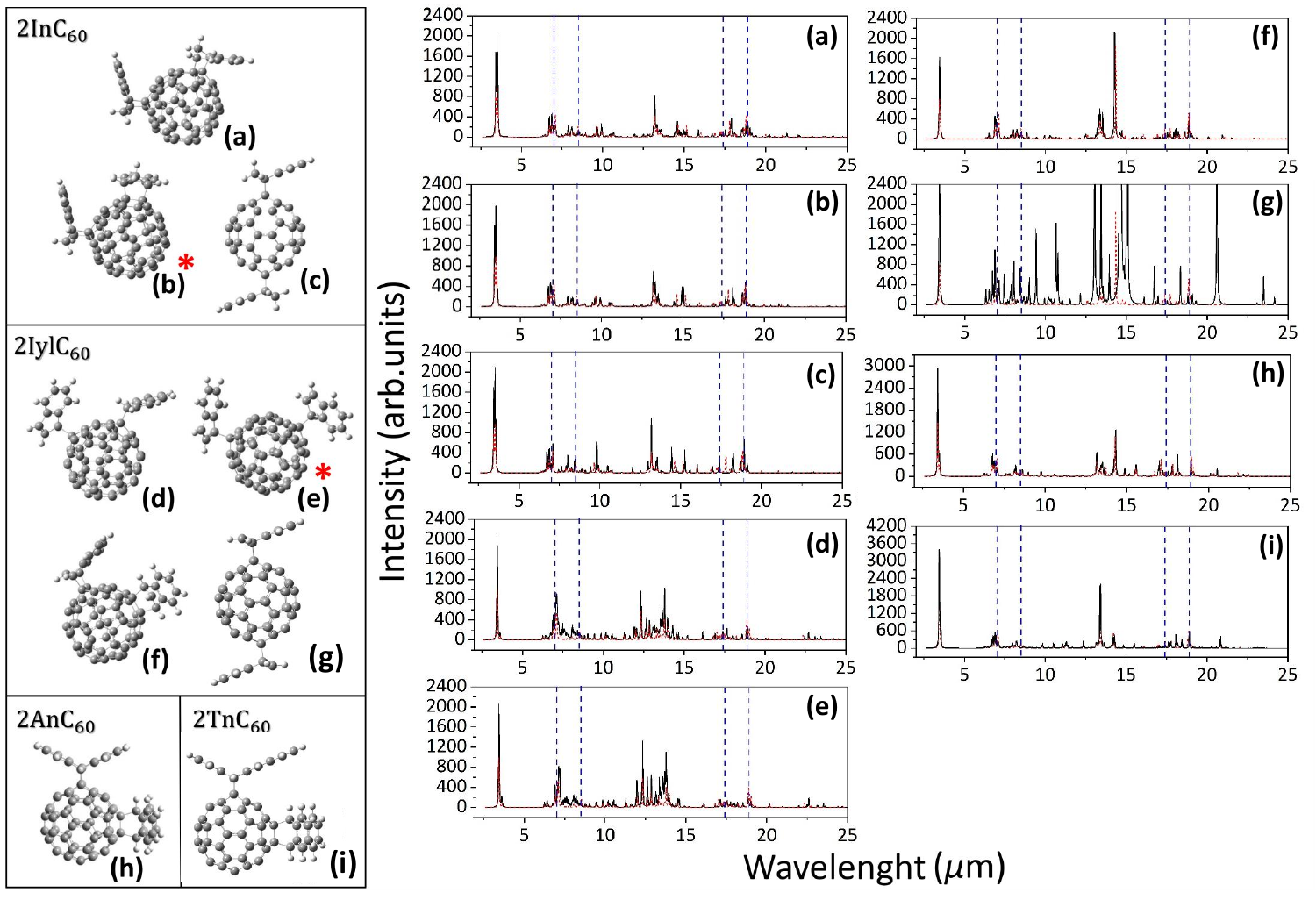}
    \caption{{\bf Left:} Models of the PAH-C$_{60}$ bis-adducts after geometry optimization for the PAHs 2H-indene (2InC$_{60}$, a-c), indenyl (2IylC$_{60}$, d-g), anthracene (2AnC$_{60}$, h), and tetracene (2TnC$_{60}$, i). In the case that the bis-adduct has different configurations (isomers), the most stable structure is highlighted with a red asterisk. {\bf Right:} The theoretical IR spectra corresponding to the bis-adducts structures a-i presented in the left panel. In all panels, the red dashed spectra correspond to the mono-adduct analogous, while the blue dashed lines mark the four strongest C$_{60}$ features ($\sim$7.0, 8.5, 17.4 and 18.9 $\mu$m). Note that the convolution parameters are the same as in Figures \ref{fig4} and \ref{fig5a}.}
    \label{figb1}
\end{figure*}
Multiple configurations (isomers) can be generated for the bis-adducts according to the arrangement of the PAHs bonded to the C$_{60}$ cage. The isomers are molecular structures defined by the same chemical formula but with different spatial distribution. The number of possible isomers (or structures) is extremely large because it depends on the degrees of freedoms for the PAHs and C$_{60}$. Thus, in order to reduce the number of possible models we have used the stability and isomerization information available in the literature. Previous theoretical works have studied the stability and isomerization of anthracene bis-adducts, allowing us to discriminate between the different models and select the most stable structures \citep{SABIROV2016,Rodrigues2022}. This isomerization information was also used to built the rest of the bis-adduct models, facilitating our theoretical predictions by significantly reducing the amount of models under consideration. Figure \ref{figb1} displays the bis-adducts structures considered together with their corresponding most stable isomers. For the anthracene (An) case, only one structure is shown; i.e. the most stable configuration as demonstrated by previous and extensive theoretical studies \citep{SABIROV2016,Rodrigues2022}. Furthermore, as already mentioned in Sect. \ref{SCF}, the anthracene bis-adduct (2AnC$_{60}$) has been obtained precisely using a novel experimental synthetic method \citep{PUJALS2022} and used for the validation of scaling factors.

The spectra displayed in Figures \ref{figb1}a-c, which correspond to the bis-adducts of 2H-indene, do not exhibit significant differences between them; the only difference is the isomer {\it c} (Figure \ref{figb1}c). Actually, it is quite difficult to distinguish the IR emission produced by these isomers of 2H-indene bis-adducts. The different configuration of the PAHs with respect to the C$_{60}$ cage neither the increment of PAH units implicate a noticeable change of the IR spectra, in comparison to the mono-adduct analogous (1InC$_{60}$$^{'}$, see Figure \ref{fig4}). The most prominent change is seen for the C-H stretching feature at $\sim$3.3-3.6 $\mu$m, which increases its intensity, reflecting the increment of CH units from mono- to bis-adducts.

On the contrary, the IR spectra produced by the isomers of indenyl bis-adducts display noticeable differences between them and also when compared to their mono-adduct analogous (Figures \ref{figb1}d-g and \ref{app2}); with the clear presence of distinctive IR spectral features depending on the isomer. The indenyl isomers have these specific features mainly due to the change of the PAH binding to the C$_{60}$ cage. The PAH can bind C$_{60}$ with one or two C-C bonds depending on the structure, while for 2H-indene the binding is always the same. Models like those in Figures \ref{figb1}d-e denote single C-C bonds between the PAH and C$_{60}$. Both exhibit subtle differences in terms of intensity, but they possess four distinctive features: (i) a broad plateau feature from 13 to 14.7 $\mu$m; (ii) a clear discrete $\sim$12.4 $\mu$m feature accompanied with well-defined peaks around it; (iii) a broad 6.7-7.5 $\mu$m band; and (iv) a strong $\sim$3.3-3.5 $\mu$m feature accompanied by a satellite feature at $\sim$3.6 $\mu$m.

A higher stability is not always an indication of more detectable features, and  the complexity of the IR spectra depends more on the charge reordering and change in the dipole moment induced by the PAH-C$_{60}$ bonding. The former is clearly understandable observing the spectra in Figures \ref{figb1}f-g, corresponding to the less stable isomers of indenyl bis-adducts. The specific features observed in these spectra are quite noticeable in terms of intensity; for instance, Figure \ref{figb1}f shows an intense narrow signal at $\sim$14.2 $\mu$m accompanied by broader feature centered at $\sim$13.3 $\mu$m. Under this configuration {\it f}, the IR spectrum shows a resemblance with its mono-adduct analogous (1IylC$_{60}$$^{'}$), maintaining almost the same features. Such behavior indicates that binding two indenyl molecules to C$_{60}$ following this geometry almost does not destroy the symmetry of the vibrations. Interestingly, the model in Figure \ref{figb1}g results in weakly bonded indenyl molecules, but it shows the richer IR spectrum of all of the bis-adduct models. Multiple features (e.g. at $\sim$13.0, 14.6, 15.0 and 20.6 $\mu$m) surpass in intensity the $\sim$3.3-3.6 $\mu$m band, which is usually the most intense signal in the PAH-C$_{60}$ adducts described so far; where most of the vibrations, implying C-C stretching and C-H out-of-plane, close to the PAH-C$_{60}$ bond create a distortion similarly to a Jahn-Teller effect \citep{Dunn2018}\footnote{In the case of the model in Figure \ref{figb1}g the point group is defined as \textit{C$_2$h}, which is one of the symmetry subgroups producing a Jahn-Teller distortion in C$_{60}$ derivatives.}. Seemingly, these C-H out-of-plane vibrations are IR active in the bis-adduct due to the higher planarity and aromaticity of the indenyl structure. This combined to the configuration in Figure \ref{figb1}g produce the strong features at $\sim$14.6, 15.0 and 20.6 $\mu$m. However, the analogous bis-adduct for 2H-indene with a similar configuration (Figure \ref{figb1}c) does not display such C-H out-of-plane active vibrations because the presence of CH$_2$ in 2H-indene reduces the planarity and aromaticity of the molecular structure. Nevertheless, it is likely that the indenyl bis-adduct (Figure \ref{figb1}g) should be a species with a rather short-lifetime as a consequence of the weak bonds between the PAHs and C$_{60}$. Finally, the IR features seen in the simulated spectra of the models for the An and Tn bis-adducts (Figures \ref{figb1}h-i) are characterized only by an increment of CH stretching (3.3-3.6 $\mu$m) due to the amount of added bonds.

\section{Tris-adducts}\label{tris}
\begin{figure}
    \centering
    \includegraphics[width = 8.7cm]{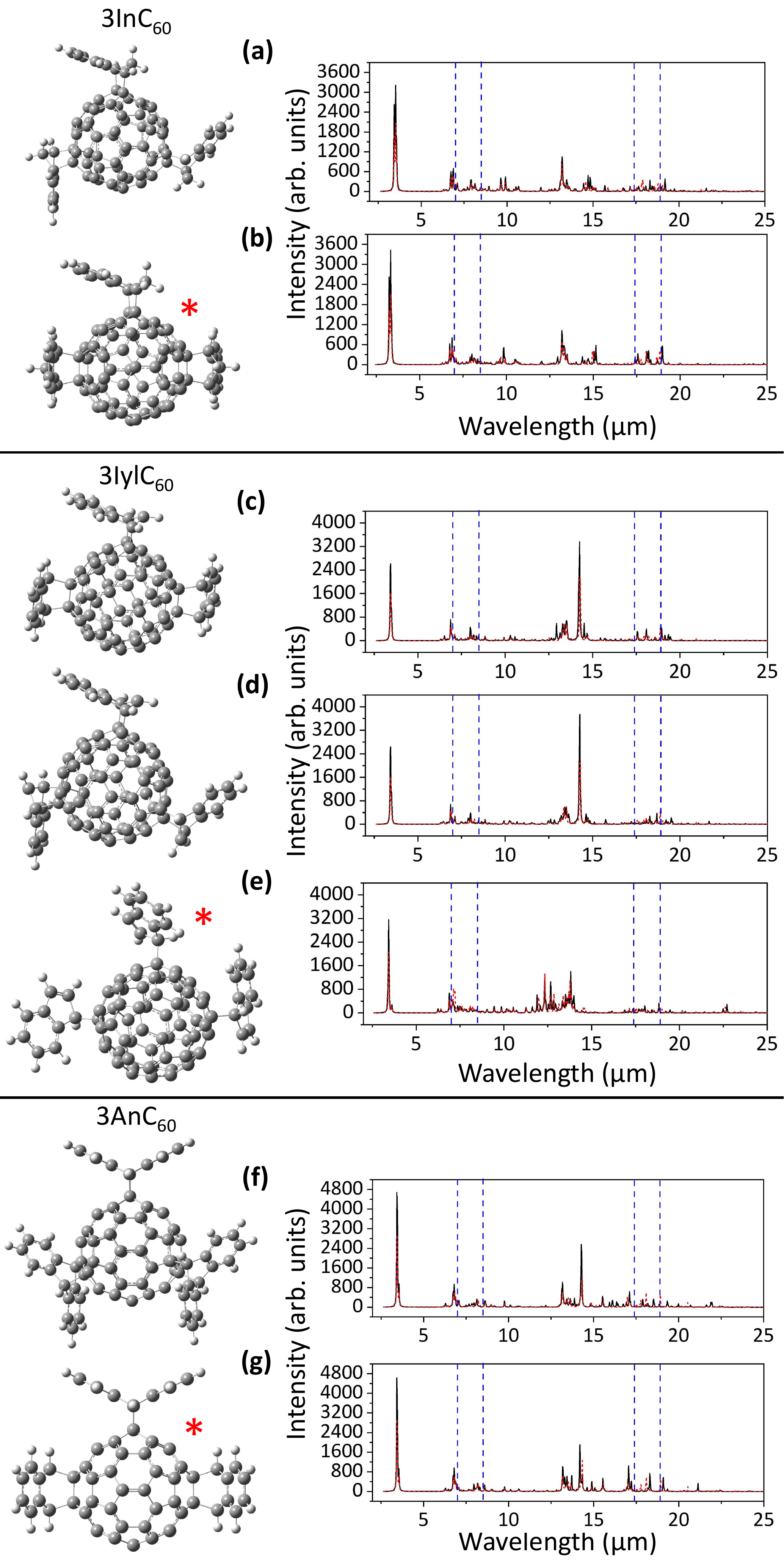}
    \caption{
  {\bf Left:} Models of the PAH-C$_{60}$ tris-adducts after geometry optimization for the PAHs 2H-indene (3InC$_{60}$, a-b), indenyl (3IylC$_{60}$, c-e) and anthracene (3AnC$_{60}$, f-g). In the case that the tris-adduct has different configurations (isomers), the most stable structure is highlighted with a red asterisk. {\bf Right:} The theoretical IR spectra corresponding to the tris-adducts structures a-g presented in the left panel. In all panels, the red dashed spectra correspond to the bis-adduct analogous, while the blue dashed lines mark the four strongest C$_{60}$ features ($\sim$7.0, 8.5, 17.4 and 18.9 $\mu$m). Note that the convolution parameters are the same as in Figures \ref{fig4} and \ref{fig5a}.}
    \label{figb2}
\end{figure}

In order to build the models of the PAH-C$_{60}$ tris-adducts, we have selected the corresponding bis-adduct from the previous Appendix \ref{bis} and added another PAH unit to the structure. Arguably, this could bias the final structures since we did not explore multiple possible isomers, but it was our preferred way to screening the number of structures; which again, would be huge due to the high number of degrees of freedom. Figure \ref{figb2} shows the models and IR spectra of the tris-adducts for In, Ilyl and An. As we have mentioned before, the C$_{60}$ tris-adducts models for tetracene and pentacene were not built because of the computational limitations.

Figure \ref{figb2} shows non-significant differences in the IR spectra of the same PAH-C$_{60}$ tris-adduct; except, again for the C$_{60}$ tris-adducts with indenyl (3IylC$_{60}$, panels c-e in Figure \ref{figb2}). In particular, for configuration {\it e} where the spectral change is due to the single C-C bond between the PAH and C$_{60}$. Furthermore, the 3IylC$_{60}$ spectra are still very similar to the ones from their mono-adducts analogous; although, the intensity varies for the most important vibrations that depend on the number of CH bonds (see also Figure \ref{fig4}). The forces exerted over the C$_{60}$ cage by the binding of three PAH units create a highly symmetric carbon cage, which is equivalent to the mono-adducts and that is reflected in the spectra similarities. Thus, according to our theoretical predictions it should be very difficult to distinguish, in terms of pure IR emission features, mono- from tris-adducts. 

In summary, the main characteristics of the C$_{60}$ tris-adducts spectra are: (i) the more noticeable broad feature, centered at $\sim$3.5 $\mu$m, for 2H-indene (Figures \ref{figb2}a-b); and (ii) the presence of a weaker red-shifted feature at $\sim$3.6 $\mu$m (Figures \ref{figb2}e-g), which denotes the increment of non-equivalent CH bonds by symmetry directly connected to C$_{60}$. Such features were difficult to distinguish in the case of mono- and bis-adducts due to the lower number of CH bonds implicated. Both features are related to CH bonds but imply different chemical environments; for 2H-indene tris-adducts, they are a consequence of the contribution of more CH$_2$ bonds, while for the indenyl and anthracene cases, they are related to the CH closest to the C-C bonded to C$_{60}$. Finally, it is worth noting that in the case of indenyl, the feature at $\sim$3.6 $\mu$m only appears in the model {\it e} with a C-C single bond to C$_{60}$ (see Figures \ref{figb2}c-e and Appendix \ref{ap2} for a comparison).

\FloatBarrier\section{Abundance estimation from kinetics data for the construction of abundance-weighted IR spectra of PAH-C$_{60}$ adducts}\label{ap}
We have estimated the expected abundances of PAH-C$_{60}$ adducts from the kinetics data previously reported. For the case of indenyl ($\rm C_9H_7$), the data extracted from indene ($\rm C_9H_8$) pyrolysis \citep{JIN2019,Lifshitz2004,POUSSE2010} has been used. Note that indene pyrolysis is also connected to the anthracene formation, through the following mechanism in the presence of cyclopentadiene ($\rm C_5H_6$) \citep{MULHOLLAND2000}:
\begin{equation*}
\rm C_9H_8+C_5H_6 \xrightarrow{-2H} C_{14}H_{12}\mkern3mu(C_9H_7-C_5H_5) \xrightarrow{-2H} C_{14}H_{10}    
\end{equation*}
Therefore, with the exception of pentacene for which not kinetic data is available, it is possible to roughly estimate the expected abundances of PAH-C$_{60}$ adducts by using the corresponding equations, knowing that they follow a second order rate law:

\begin{equation} \label{a2}
   \rm  [A] = \frac{[A]_0}{1+kt\cdot[A]_0} 
\end{equation}
where k is the rate constant, [A] is the concentration of the reactant at a given time t and [A]$_0$ is the initial concentration of the reactant. In order to apply Equation \ref{a2} we have assumed the following assumptions:

\begin{itemize}
    \item The PAH-C$_{60}$ abundances depend only on the PAH abundance.
    \item The abundances of the Iyl-C$_{60}$ and An-C$_{60}$ adducts depend only on the indene abundance, following the pyrolysis mechanism mentioned above.
    \item The initial abundance [A]$_0$ of indene is obtained from the column density estimates towards the cold dark cloud TMC-1 reported by \citet{Cernicharo2021} and \citet{Burkhardt2021}, which are 1.6 $\times$ 10$^{13}$ and 0.96 $\times$ 10$^{13}$ cm$^{-2}$, respectively\footnote{The column density is transformed to abundance (or concentration, in mol/cm$^{3}$) by assuming the TMC-1 geometrical model of \citet{avery1982} and a distance of 140 pc; i.e. basically along a path of 0.14 pc.}.
    \item The transition from mono- to bis-adducts follows the same kinetics as from bis- to tris-adducts, using as reference the solid state kinetics of InC$_{60}$ by \citet{RODRIGUES2023}.
\end{itemize}
According to the above-mentioned principles we have determined the PAH abundances of indenyl, anthracene and tetracene from indene in order to estimate the corresponding PAH-C$_{60}$ adducts abundances. For this purpose, we refer again to Equation \ref{a2}, where it can be noticed that the timescale (t) has to be introduced. This timescale can be first computed from the indene pyrolisis \citep{MULHOLLAND2000} using the half-life time expression:

\begin{equation}
\rm t_{1/2} = \frac{1}{k[A]_0} \label{a3}    
\end{equation}

we remind that in Equation \ref{a3} [A]$_0$ is the indene abundance from \citet{Cernicharo2021} and \citet{Burkhardt2021}. Using the value in $\rm t_{1/2}$ and the experimental constant rates (k) of the PAH-C$_{60}$ adducts we can determine [A] and [A]$_0$ for indene and anthracene. For indenyl, we directly used the result obtained from the indene pyrolisis, while for the case of the tetracene adducts we have used the relation of the constant rates with anthracene \citep[see Tables 1 and 2 in][]{SAROVA2004}:

\begin{equation}
\rm k_{TnC_{60}} = 1.94\cdot 10^{2}k_{AnC_{60}} \label{a1}    
\end{equation}

It should be noted that [A] and [A]$_0$ the reactants, and thus the abundances of products must be derived from [A]$_{prod}$=[A]$_0$-[A], which describes how much of the reactant has become a product. Table \ref{tb_abundance} displays the estimated expected abundance for each PAH-C$_{60}$ adduct under study. Clearly, from the listed values, the abundances of anthracene and tetracene are negligible compared to the ones of indene and indenyl.

\begin{table}
\centering
      \caption[]{Estimated abundances (concentrations) for the different PAH-C$_{60}$ adducts.}
      \label{tb_abundance}
         \setlength\tabcolsep{3pt}
         \renewcommand{\arraystretch}{1.2}
         
         \begin{tabular}{llcc}
            \hline
            \noalign{\smallskip}
           & PAH-C$_{60}$ Adduct & [A]$_{prod}^{\rm Cer}$  & [A]$_{prod}^{\rm Buk}$ \\ \noalign{\smallskip}
               &   & mol/cm$^3$ & mol/cm$^3$ \\ \hline \noalign{\smallskip}
\multirow{4}{*}{Mono}
&  2H-indene (In)    &  6.15$\cdot10^{-29}$   &  3.69$\cdot10^{-29}$ \\    
&  indenyl (Iyl)   &   3.07$\cdot10^{-29}$   &  1.84$\cdot10^{-29}$\\ 
&  anthracene (An)   &  2.60 $\cdot10^{-32}$   &   3.11 $\cdot10^{-32}$\\ 
&  tetracene (Tn)    &  1.00 $\cdot10^{-37}$   &   1.00 $\cdot10^{-37}$\\        \arrayrulecolor{gray} \cline{2-4} \noalign{\smallskip}
\multirow{4}{*}{Bis}
&  2H-indene (In)    &  3.55$\cdot10^{-29}$   &  2.13$\cdot10^{-29}$ \\    
&  indenyl (Iyl)   &   1.30$\cdot10^{-29}$   &  7.78$\cdot10^{-30}$\\ 
&  anthracene (An)  & 1.10$\cdot10^{-32}$     &  1.31$\cdot10^{-32}$\\ 
&  tetracene (Tn)        &  4.22$\cdot10^{-38}$     &  4.22$\cdot10^{-38}$\\\noalign{\smallskip}
\arrayrulecolor{gray} \cline{2-4} \noalign{\smallskip}
\multirow{3}{*}{Tris}
&  2H-indene (In)    &  3.55$\cdot10^{-29}$   &  2.13$\cdot10^{-29}$ \\    
&  indenyl (Iyl)   &   1.30$\cdot10^{-29}$   &  7.78$\cdot10^{-30}$\\ 
&  anthracene (An)        &   1.10$\cdot10^{-32}$  &   1.31$\cdot10^{-32}$\\ \arrayrulecolor{black}\noalign{\smallskip}\hline 
         \end{tabular}
      \tablefoot{[A]$_{prod}^{\rm Cer}$ and [A]$_{prod}^{\rm Buk}$ correspond to the abundances determined using the initial abundance [A]$_0$ of indene as estimated from the column density values by \citet{Cernicharo2021} and \citet{Burkhardt2021}, respectively (see the text for more details).}
   \end{table}

On the other hand, we also consider the probability of formation according to the thermodynamic stability of PAH-C$_{60}$ adducts and its influence on the IR spectra. For this, a Boltzmann distribution (equation \ref{a4} below) has been used to weight the different regioisomers for the bis- and tris-adducts (see Table \ref{tb2}), which is described as:
\begin{equation}
  \rm   P_i=\frac{e^{-\Delta H^i_f/kT}}{\displaystyle\sum^M_{j=1}e^{-\Delta H^j_f/kT}}\label{a4}
\end{equation}
with $\rm \Delta H^i_f$ being the Enthalpy of formation of the bis- or tris-adduct, M the maximum number of regioisomers, T the temperature and k the Boltzmann constant. Note that $\rm P_i=1$ for the 2AnC$_{60}$ and 2TnC$_{60}$ adducts because no regioisomers were considered. We also note that $\rm \Delta G_f=\Delta H_f - T\Delta S_f$ is the expression that connects these $\rm \Delta H_f$ values with the $\rm \Delta G_f$ ones in Table \ref{tb2}; both were calculated at 300 K and are directly extracted from our quantum-chemistry calculations. By using the quantities obtained so far, the IR intensity for each PAH-C$_{60}$ adduct is thus computed by the following expression:

\begin{equation}
  \rm I \propto P_i\cdot[A]_{\mathit{prod}}\cdot\frac{\partial\mu}{\partial x} 
\end{equation}
 It is important to note here that $\partial\mu/\partial x$ it is the change in dipole moment ($\partial\mu$) as a function of the displacement produced by the vibration or vibrational mode ($\partial x$). This $\partial\mu/\partial x$ is the IR cross section as determined from our quantum-chemistry calculations; i.e. without convolving with any peak profile function (Lorentzian, Gaussian, etc.). Finally, the abundance-weighted PAH-C$_{60}$ mixture spectra (displayed in Figures \ref{fig8}b and c) were constructed by summing the individual weighted PAH-C$_{60}$ adduct IR spectra; as obtained for the two initial abundances [A]$_0$ of indene from \citet{Cernicharo2021} and \citet{Burkhardt2021}. Evidently, due to $[A]_{prod}$ and P$\rm _i$ PAH-C$_{60}$ adducts have a lower/higher contribution to the abundance-weighted spectra.

\section{Details on the 1TnC$_{60}$ mono-adduct configuration}\label{ap1}
\begin{figure}[h!]
    \centering
    \includegraphics[width = 8.5cm]{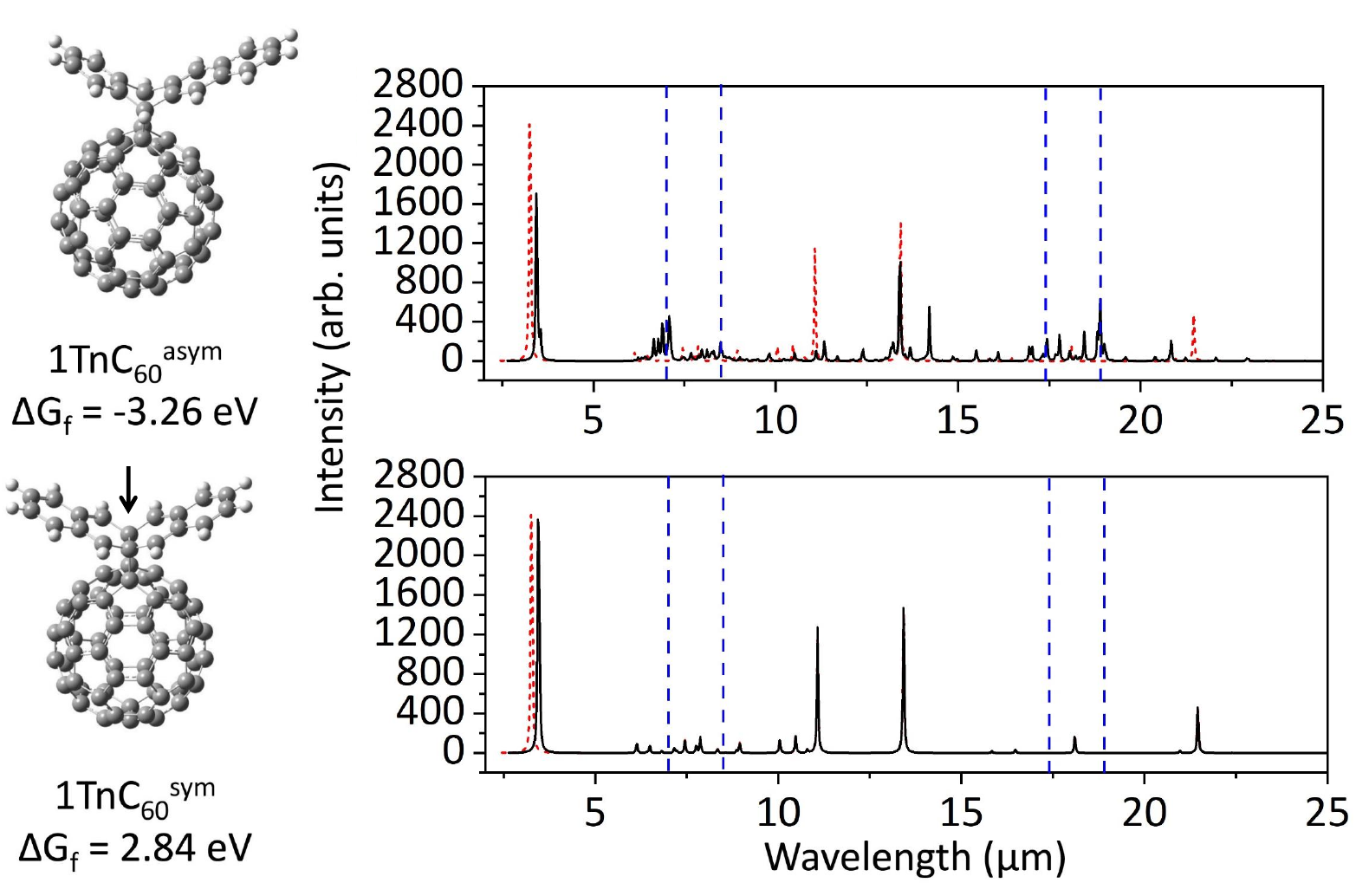}
    \caption{\textbf{Left:} Representation of the 1TnC$_{60}$ mono-adduct models in their asymmetric (1TnC$_{60}^{\rm asym}$) and symmetric (1TnC$_{60}^{\rm sym}$) binding configurations. The corresponding Gibbs free energies of formation are indicated. \textbf{Right:} The theoretical IR spectra for each 1TnC$_{60}$ mono-adduct configuration.}
    \label{figap}
\end{figure}

Structurally, it is possible to build two mono-adduct configurations for 1TnC$_{60}$, bonding the PAH by a symmetric or asymmetric attachment (see Figure \ref{figap}). The symmetric model represented in Figure \ref{figap} is much more thermodynamically unstable than its asymmetric counterpart. Since we have established the possible formation of 1TnC$_{60}$ through an exohedral cycloaddition, the binding to C$_{60}$ must occur inside an hexagon ring instead on the C-C bond bridging the fused rings, which corresponds to the asymmetric model. In addition, the asymmetric model is in agreement with previous experimental studies that suggest that tetracene binds C$_{60}$ in such a way \citep{SAROVA2004}. 
\section{High-resolution windows of PAH-C$_{60}$ adduct spectra}\label{ap2}
\begin{figure*}
    \centering
    \includegraphics[width = 15.5cm]{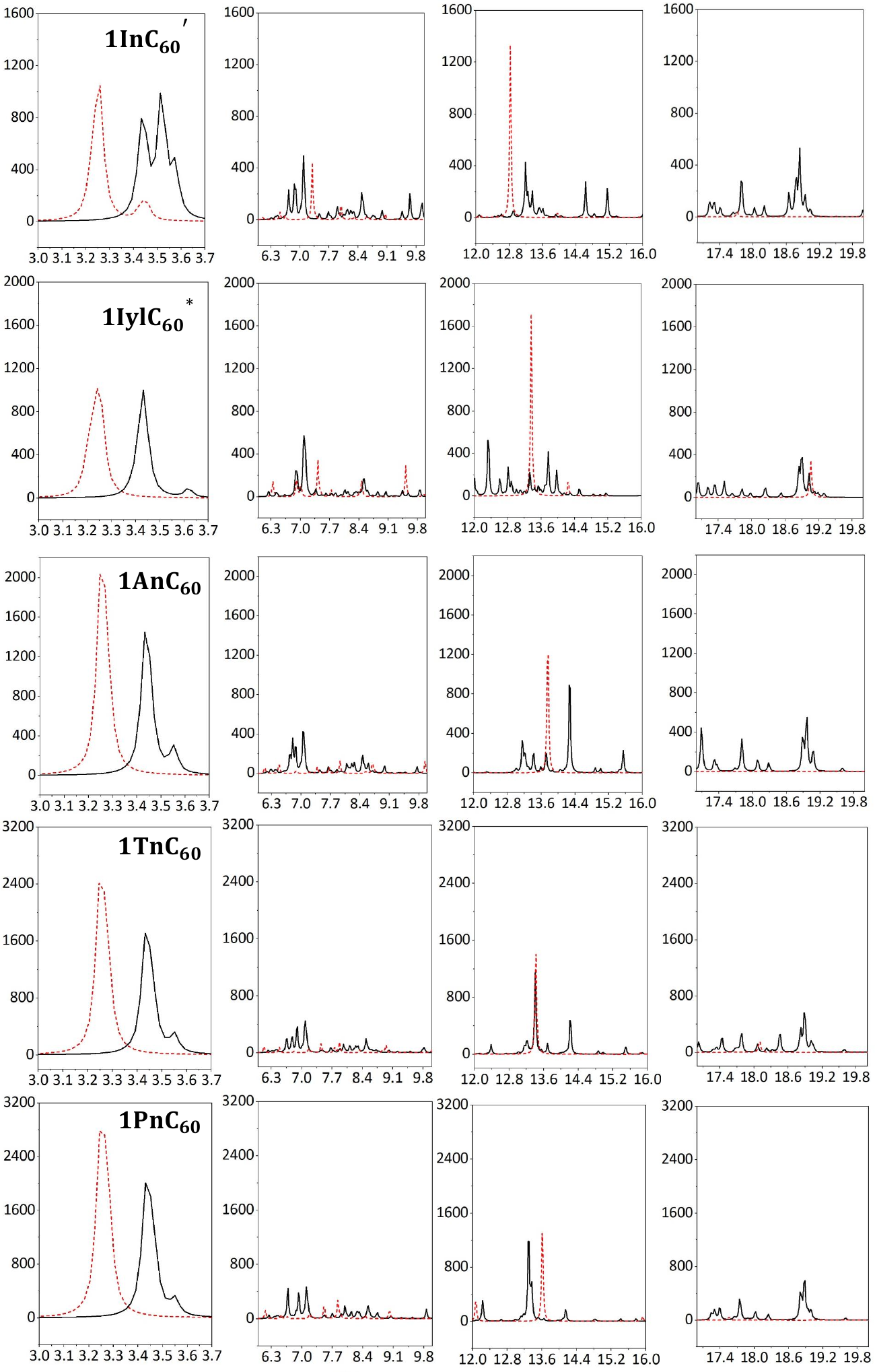}
    \caption{High-resolution windows of the more stable mono-adducts presented in Figure \ref{fig4}. Red dotted lines represent the free PAHs.}
    \label{app1}
\end{figure*}
\begin{figure*}
    \centering
    \includegraphics[width = 15.5cm]{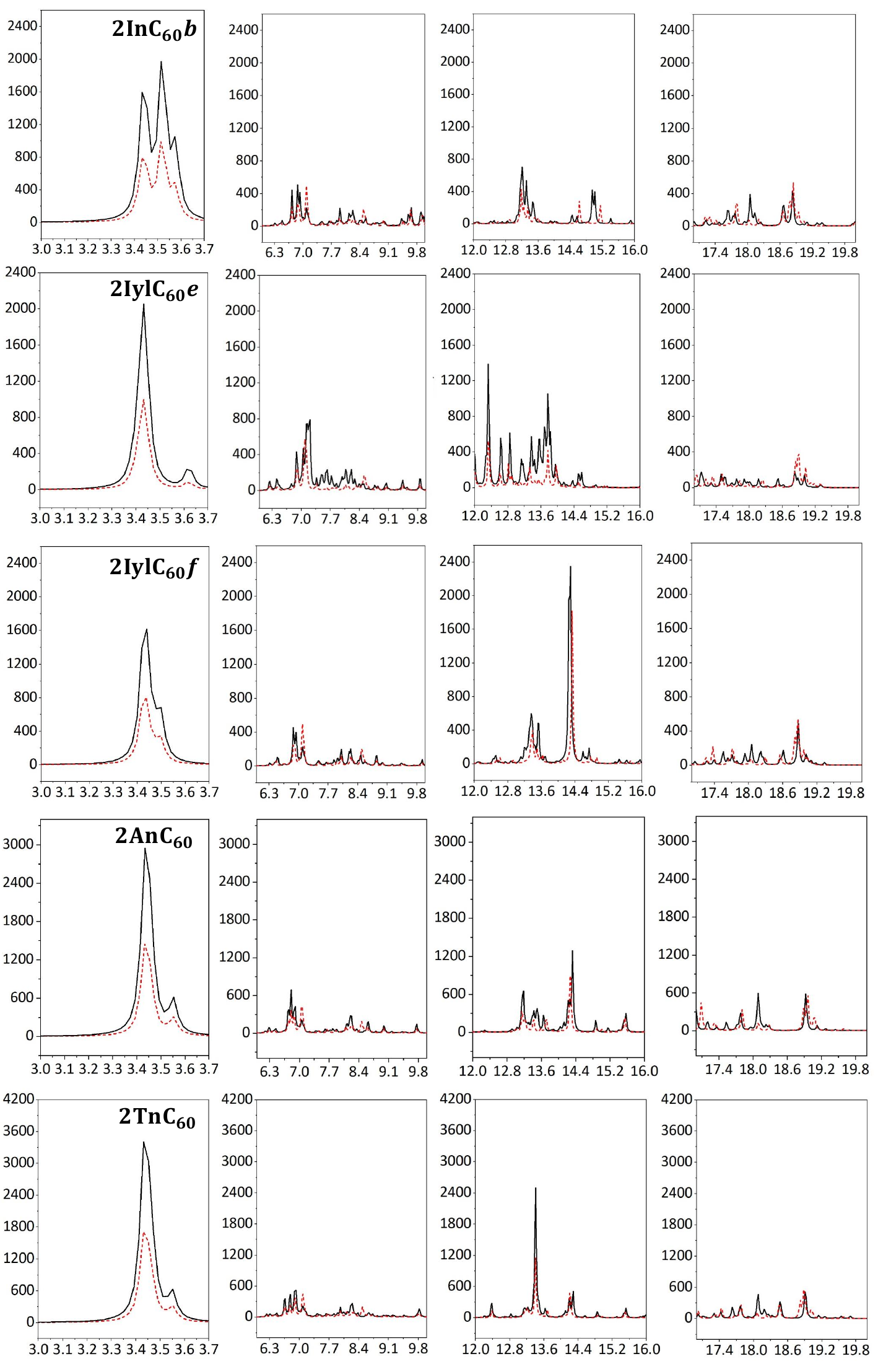}
    \caption{High-resolution windows of the more stable bis-adducts presented in Figure \ref{figb1}.}
    \label{app2}
\end{figure*}
\begin{figure*}
    \centering
    \includegraphics[width = 15.5cm]{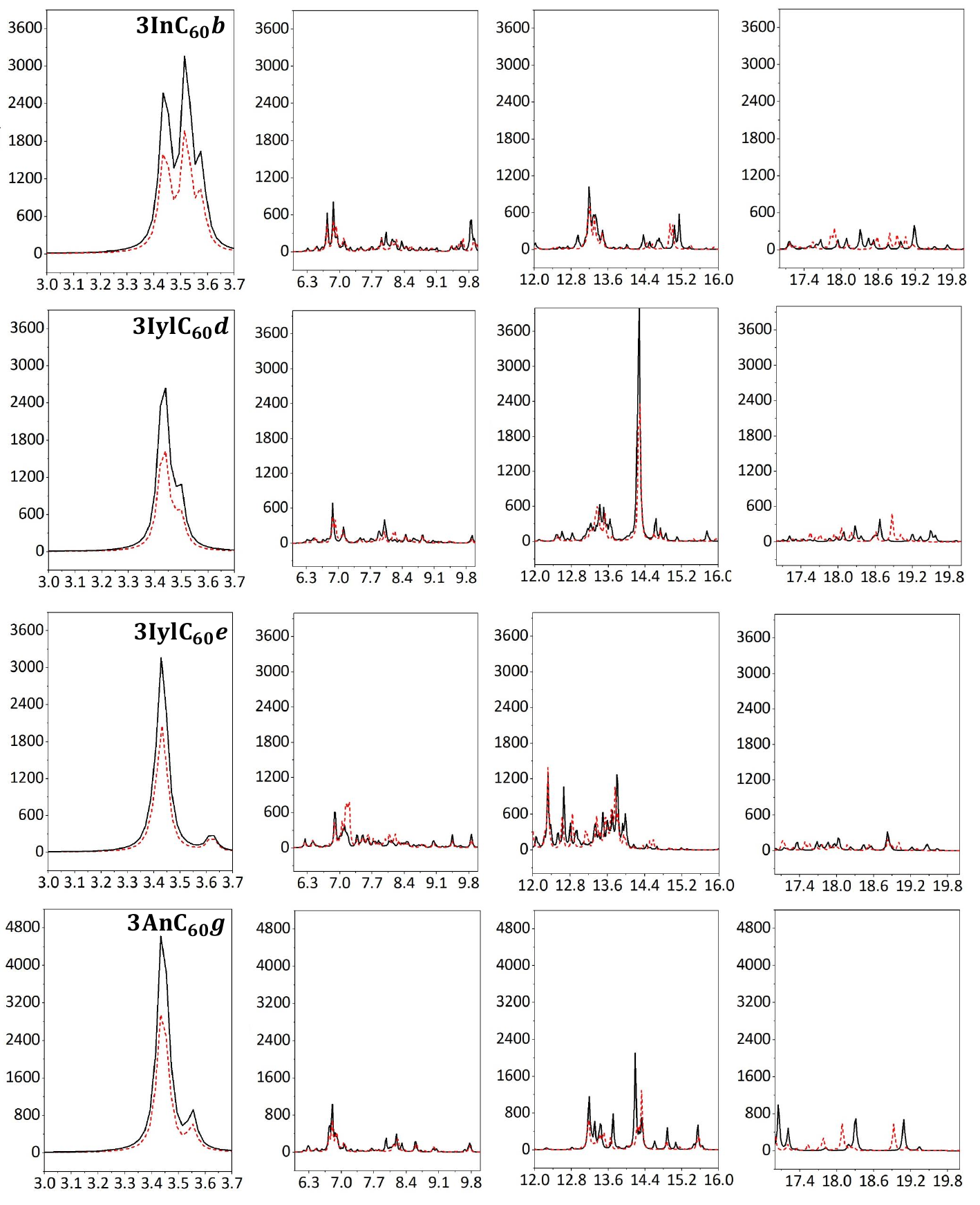}
    \caption{High-resolution windows of the more stable tris-adducts presented in Figure \ref{figb2}.}
    \label{app3}
\end{figure*}
\begin{figure*}
    \centering
    \includegraphics[width = 15.5cm]{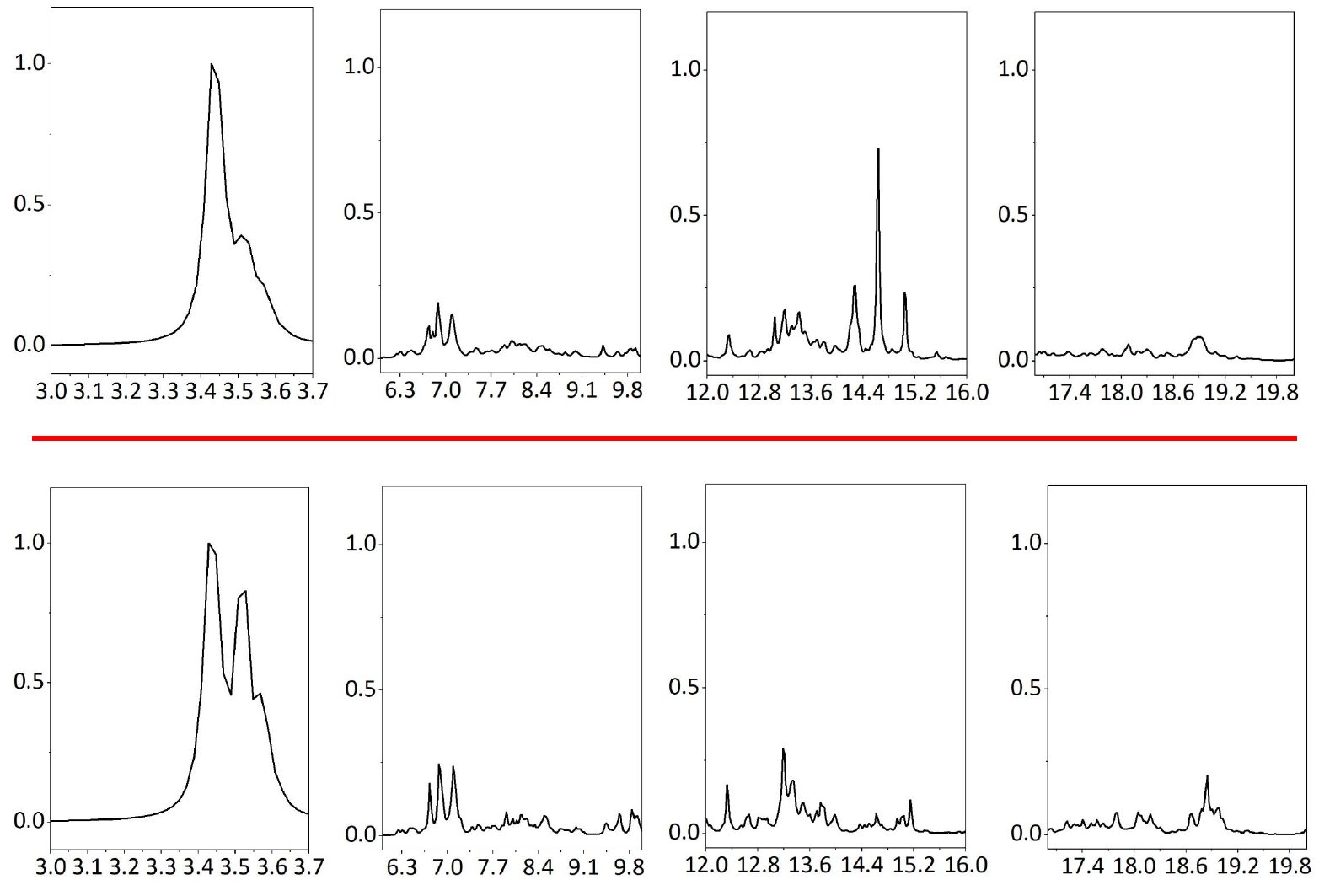}
    \caption{High-resolution windows of the summed quantum-chemical spectra (top) and the abundance-weighted spectra (bottom) presented in Figure \ref{fig8}.}
    \label{app4}
\end{figure*}

\end{appendix}
\end{document}